\documentclass[]{aastex7}
\usepackage{subcaption}
\usepackage{multirow}
\usepackage{amsmath}
\usepackage{comment}
\usepackage{booktabs}
\usepackage{amssymb}
\let\longtable*\relax
\begin{document}

\title{Reinterpreting the JWST Observations of 55 Cancri e with a Non-Grey General Circulation Model}

\author{Ruizhi Zhan}
\affiliation{Dept.\ of Atmospheric and Oceanic Sciences, School of Physics, Peking University, Beijing 100871, People’s Republic of China}
\email{ruizhi.zhan@stu.pku.edu.cn}
\author{Daniel D.B. Koll}
\affiliation{Dept.\ of Atmospheric and Oceanic Sciences, School of Physics, Peking University, Beijing 100871, People’s Republic of China}
\email{dkoll@pku.edu.cn}
\correspondingauthor{Daniel D.B. Koll}
\accepted{for publication in ApJ}

\begin{abstract}
Recent observations of 55 Cancri e suggest an atmosphere rich in CO or CO$_2$ \citep{hu_secondary_2024}; other observations indicate the planet's eclipse depth is highly variable \citep[e.g.][]{patel_jwst_2024}. So far, these observations have only been interpreted using 1D models without self-consistent heat redistribution, as the planet's extreme temperatures make it inaccessible to most 3D models. Here we perform cloud-free GCM simulations of 55 Cancri e using custom correlated-$k$ coefficients developed from the ExoMol database. Our best-fit simulations match the JWST spectra from \citet{hu_secondary_2024} well, favoring an atmosphere that is both thick ($\ge$ 10 bar) and CO$_2$-rich ($>1\%$ CO$_2$ volume mixing ratio), while ruling out thin ($<$ 10 bar) and pure-CO/CO$_2$-poor atmosphere, which were previously proposed based on 1D models \citep{hu_secondary_2024,zilinskas_characterising_2025}. We also find large-scale atmospheric dynamics, i.e. weather, is insufficient to explain the observed variability. A thick, CO$_2$-rich atmosphere implies that 55 Cancri e likely formed with significantly more volatiles than Earth and Venus. In addition, a thick atmosphere makes it unlikely that the planet's variability is caused by transient outgassing \citep{heng_transient_2023}, favoring other variability mechanisms (e.g. clouds). Our work provides model constraints for upcoming JWST observations of 55 Cancri e, and highlights the importance of interpreting thermal emission observations with self-consistent 3D models. 
\end{abstract}


\section{Introduction}\label{sec::introduction}

Characterizing the atmospheres of rocky exoplanets promises to provide unique insight into how these planets form and evolve. One prime candidate is the super-Earth 55 Cancri e. Orbiting a bright K star, with an ultra-short-period orbit of $\sim18$ h, an equilibrium temperature of $\sim$2000 K, and a density lower than expected for an Earth-like bulk composition, 55 Cancri e is one of the most accessible targets for atmospheric characterization with current and future space telescopes \citep[e.g.][]{demory_map_2016,hu_secondary_2024,patel_jwst_2024}.

Pre-JWST observations suggested 55 Cancri e might support an atmosphere, but its existence and composition remained highly debated. Early Spitzer phase curve observations found a significant hot spot offset and efficient heat redistribution, indicating an atmosphere \citep{demory_map_2016, angelo_case_2017}. However, these conclusions were challenged by later reanalysis \citep{mercier_revisiting_2022}. Transit spectroscopy from HST and ground-based facilities were able to rule out a low mean molecular weight (MMW) atmosphere, but remained inconclusive about the planet having a high MMW atmosphere or no atmosphere at all \citep{tsiaras_detection_2016,jindal_characterization_2020, deibert_near-infrared_2021}.

The most compelling evidence for an atmosphere on 55 Cancri e was recently obtained via JWST eclipse spectra with NIRCam and MIRI \citep{hu_secondary_2024}. Interpreting the data using 1D models, \citet{hu_secondary_2024} found robust evidence for an atmosphere, composed of CO or CO$_2$ with N$_2$ as a potential background gas. 
The same study also ruled out an atmosphere made of silicate-dominated rock vapor. This surprisingly suggests that 55 Cancri e was able to retain a volatile-rich atmosphere against atmospheric erosion, which is expected to be extremely efficient on such a hot and close-in planet.
A follow-up study reanalysed the same spectra using a larger grid of 1D models and found the data are compatible with an N$_2$-CO$_2$ atmosphere, but can also be matched by alternative compositions such as CO, H$_2$O, PO, or PH$_3$ \citep{zilinskas_characterising_2025}.

Other observations are compatible with 55 Cancri e having an atmosphere, but show that the planet's eclipse depth varies strongly over time. Recent JWST observations found the planet's brightness temperature at 4.5 $\mu$m varies by at least a factor of 3 on sub-weekly timescales \citep{patel_jwst_2024}. Similar temporal variability was also found in older Spitzer observations \citep{demory_variability_2016}, and in TESS and CHEOPS data at shorter wavelengths \citep{valdes_weak_2022, demory_55_2023}.

So far, the observed time variability remains poorly understood. Proposed explanations include silicate cloud oscillations driven by a magma–cloud feedback \citep{loftus_extreme_2024}, tidal heating waves \citep{farhat_magma_2026}, and stochastic volcanism that generates a transient thin atmosphere which then undergoes atmospheric escape \citep{heng_transient_2023}. However, the variability could also be external to the planet, for example due to an inhomogeneous circumstellar dust torus \citep{patel_jwst_2024}.

One major shortcoming in the current discussion is that the best data of 55 Cancri e are all thermal emission observations, which are strongly sensitive to an atmosphere's horizontal heat redistribution and thus its 3D dynamics \citep[e.g.,][]{seager_method_2009,selsis_thermal_2011,koll_deciphering_2015}. However, so far the JWST data have only been interpreted using simplified 1D models, which must parameterize the atmosphere's heat redistribution \citep{hu_secondary_2024,zilinskas_characterising_2025}. This is typically done by adjusting the planet’s dayside energy budget via an efficiency factor $f$ \citep{seager_dayside_2005},
\begin{equation}
    f=\frac{1}{4}\left( \frac{T_\mathrm{day}}{T_\mathrm{eq}} \right)^4,
    \label{eq::brightness_temp}
\end{equation}
where $T_\mathrm{day}$ is the observed dayside brightness temperature, and $T_\mathrm{eq}$ is the equilibrium temperature of a isotropic blackbody, defined as $T_{\mathrm{eq}}=T_*\left( \frac{R_*}{a} \right)^{\frac{1}{2}}\left( \frac{1-A_B}{4}\right)^{\frac{1}{4}}.$ Here $T_*$ and $R_*$ are the star's effective temperature and radius, $a$ is the orbit's semi-major axis, and $A_B$ is the planet's bond albedo.

3D GCMs can improve on these interpretations by self-consistently simulating the atmosphere's heat redistribution. In a first study, \citet{hammond_linking_2017} simulated 55 Cancri e's atmosphere using a GCM with idealized grey radiation, and argued that the Spitzer observations existing at that time indicated the presence of a relatively thick and low MMW atmosphere. However, since \citet{hammond_linking_2017} there have not been any more GCM studies of 55 Cancri e, including any with more realistic radiative transfer. This is because the radiative transfer codes used in most GCMs lack comprehensive high-temperature absorption data, which makes then invalid at the extreme temperatures found on 55 Cancri e.

In this work, we revisit 55 Cancri e using non-grey 3D GCM simulations with customized high-temperature correlated-$k$ coefficients. Our primary goal is to reinterpret the JWST eclipse data of \citet{hu_secondary_2024}, who argued for a CO$_2$ or CO-rich atmosphere but were not able to constrain the atmosphere's thickness. A secondary goal is to investigate whether atmospheric dynamics and weather can explain the variability reported by \citet{patel_jwst_2024} and others.

\section{Methods}\label{sec:method}

\subsection{Overview}

\begin{deluxetable}{lcc}
    \tablewidth{0pt}
    \tablecaption{Planetary System Parameters \label{table:para}}
    \tablehead{
    \colhead{Parameter} & \colhead{Symbol} & \colhead{Value}
    }
    \startdata
    \multicolumn{3}{l}{Stellar Parameters} \\
    \hline
    Mass & $M_*$ & 0.905 $M_\odot$ \\
    Radius & $R_*$ & 0.943 $R_\odot$ \\
    Effective Temperature & $T_{*,\mathrm{eff}}$ & 5172 K \\
    Metallicity & [Fe/H] & 0.35 \\
    \cutinhead{Planetary Parameters}
    Mass & $M_p$ & 7.99 $M_\oplus$ \\
    Radius & $R_p$ & 1.875 $R_\oplus$ \\
    Orbital Period & $P$ & 0.7365 days \\
    Eccentricity & $e$ & 0.0 (tidally locked) \\
    \enddata
\end{deluxetable}

Our simulations use the spectral dynamical core of \texttt{Isca}\footnote{\texttt{Isca} on GitHub: execlim.github.io/Isca}, a widely used planetary GCM \citep{vallis_isca_2018,penn_atmospheric_2018,thomson_hierarchical_2019,thomson_effects_2019,lewis_dependence_2021}. \texttt{Isca} supports non-grey radiative transfer via the correlated-$k$ method \citep{fu_correlated_1992} using \texttt{SOCRATES}\footnote{\texttt{SOCRATES} from the UK Met Office: code.metoffice.gov.uk/trac/socrates}. When combined with appropriate correlated-$k$ tables, \texttt{SOCRATES} can simulate atmospheric radiative transfer from Earth to hot Jupiters \citep[e.g.,][]{amundsen_uk_2016,way_resolving_2017,christie_impact_2021,guzewich_3d_2021}.

Our main innovation here is to develop customized high-temperature correlated-k tables appropriate for 55 Cancri e and to use them with \texttt{Isca} and \texttt{SOCRATES}. The tables are computed from ExoMol\footnote{Website of ExoMol: \url{www.exomol.com}} line lists using \texttt{ExoCross}\footnote{ExoCross on GitHub: \url{https://github.com/Trovemaster/exocross}} \citep{yurchenko_exocross_2018}, and are supplemented by UV and collision-induced absorption (CIA) opacity data from HITRAN\footnote{Website of HITRAN: \url{https://hitran.org}} and the MPI-Mainz UV/VIS Spectral Atlas\footnote{Website of The MPI-Mainz UV/VIS Spectral Atlas: \url{https://uv-vis-spectral-atlas-mainz.org/uvvis/cross\_sections}}. Gas overlap absorption is treated using the equivalent extinction method \citep{amundsen_treatment_2017,turbet_trappist-1_2022}. The default spectral files are configured with 85 spectral bands, and detailed descriptions of the molecular line lists, cross-section calculations, and convergence tests used to determine this configuration are provided in Appendix \ref{appendix::configuration}.

Motivated by recent observations \citep{hu_secondary_2024} and theoretical outgassing calculations  \citep{fegley_volatile_2020}, we consider three types of atmosphere: CO$_2$-CO atmospheres with surface pressures $10^{-2}$--$10^2$ bar and CO$_2$ volume mixing ratios (VMRs) $10^0$--$10^{-8}$, plus pure CO; CO$_2$-N$_2$ atmospheres with surface pressures $10^{-2}$--$10^2$ bar and CO$_2$ VMRs $10^0$--$10^{-4}$; and pure H$_2$O atmospheres with surface pressures $10^{-2}$--$10$ bar. The planetary system parameters are from \citet{bourrier_55_2018}, shown in Table \ref{table:para}.

\subsection{GCM details}
\label{sec::method_gcm}
\texttt{Isca} is configured as follows. Based on sensitivity tests, we use `T21' resolution for H$_2$O atmospheres and `T42' resolution otherwise (Appendix \ref{appendix::horizontal_res}). We employ 40 uneven vertical levels with a default model top of $0.0081 p_s$, where $p_s$ is the surface pressure. To capture the emission level in optically thick atmospheres, the model top is raised to 10 Pa for pure H$_2$O atmospheres and 1000 Pa for CO$_2$-rich (VMR $\ge 10\%$) atmospheres. The time step is adjusted on a case-by-case basis and ranges from 0.01 to 40 s. For the most unstable simulations with $p_s \leq 0.1$ bar, we additionally add a Rayleigh drag sponge layer at $p \le 0.1 p_s$. 
The surface assumes a 0.5 m mixed-layer slab ocean with zero albedo. The surface albedo is consistent with the expected low albedo of molten silicate \citep{essack_low-albedo_2020}, TESS/CHEOPS observations in visible light \citep{kipping_detection_2020,demory_55_2023}, and previous GCM simulations of 55 Cancri e \citep{hammond_linking_2017}. 
In addition to the dynamical core, we include a dry convective adjustment scheme, but neglect internal heating, chemistry, and phase changes.

Simulations are considered equilibrated once TOA energy and dayside/nightside mean surface temperature fluctuations remain $<1\%$ over 10 orbits. Note that pure CO$_2$ cases with $p_s \leq 0.1$ bar, and mixed CO$_2$-CO/CO$_2$-N$_2$ cases with $p_s=0.01$ bar and CO$_2$ VMR $\geq 1\%$ do not converge. Simulations are flagged as collapsed if the time-averaged surface temperature falls low enough for condensation, which is determined by the local partial pressure.

For the \texttt{SOCRATES} radiative transfer we use a two-stream direct solver via the \citet{elsasser_heat_1942} scheme (thermal radiation) and the \citet{zdunkowski_investigation_1980} scheme (stellar radiation). The stellar spectrum is generated from PHOENIX stellar models using the \texttt{pysynphot} package\footnote{pysynphot.readthedocs.io} \citep{allard_phoenix_2016}.

\subsection{Comparison with JWST data}

We use the JWST NIRCam and MIRI data from \citet{hu_secondary_2024}, where the NIRCam data was reprocessed by \citet{zilinskas_characterising_2025} using an updated pipeline.
Following \citet{hu_secondary_2024}, we exclude the fully saturated MIRI points at 5.0–6.0 $\mu$m. Due to strong correlated noise in NIRCam, we also treat the vertical offset of the NIRCam eclipse depths as a free parameter. When comparing models and data, the NIRCam data is shifted vertically to minimize $\chi^2$, with a different offset for each model-data comparison.

To generate observed emission spectra $F_p/F_*$ from the equilibrated simulations, $F_p$ is computed from the observer-projected outgoing thermal flux at the top of the atmosphere. For a more accurate comparison with JWST data, we set $F_*$ to an empirical host-star spectrum derived from observations (0.8–5 $\mu$m: \citet{crossfield_acme_2012}; 5–12 $\mu$m: \citet{hu_secondary_2024}). However, since the empirical spectrum has limited wavelength coverage at short wavelengths, the GCMs are run using the PHOENIX stellar model. To derive the high-resolution spectra in Figure \ref{fig::eclip_spec}, we rerun the GCM for a single timestep with a higher-resolution spectral file ($R \sim 1000$, 1529 bands) applied to the equilibrated temperature profiles.

\section{Results} 
\label{sec:result}
\subsection{3D General Circulation Model Simulations of 55 Cancri e}
\label{sec:result_gcm}

\begin{figure}[htbp]
    \centering
    \includegraphics[width=1.0\textwidth]{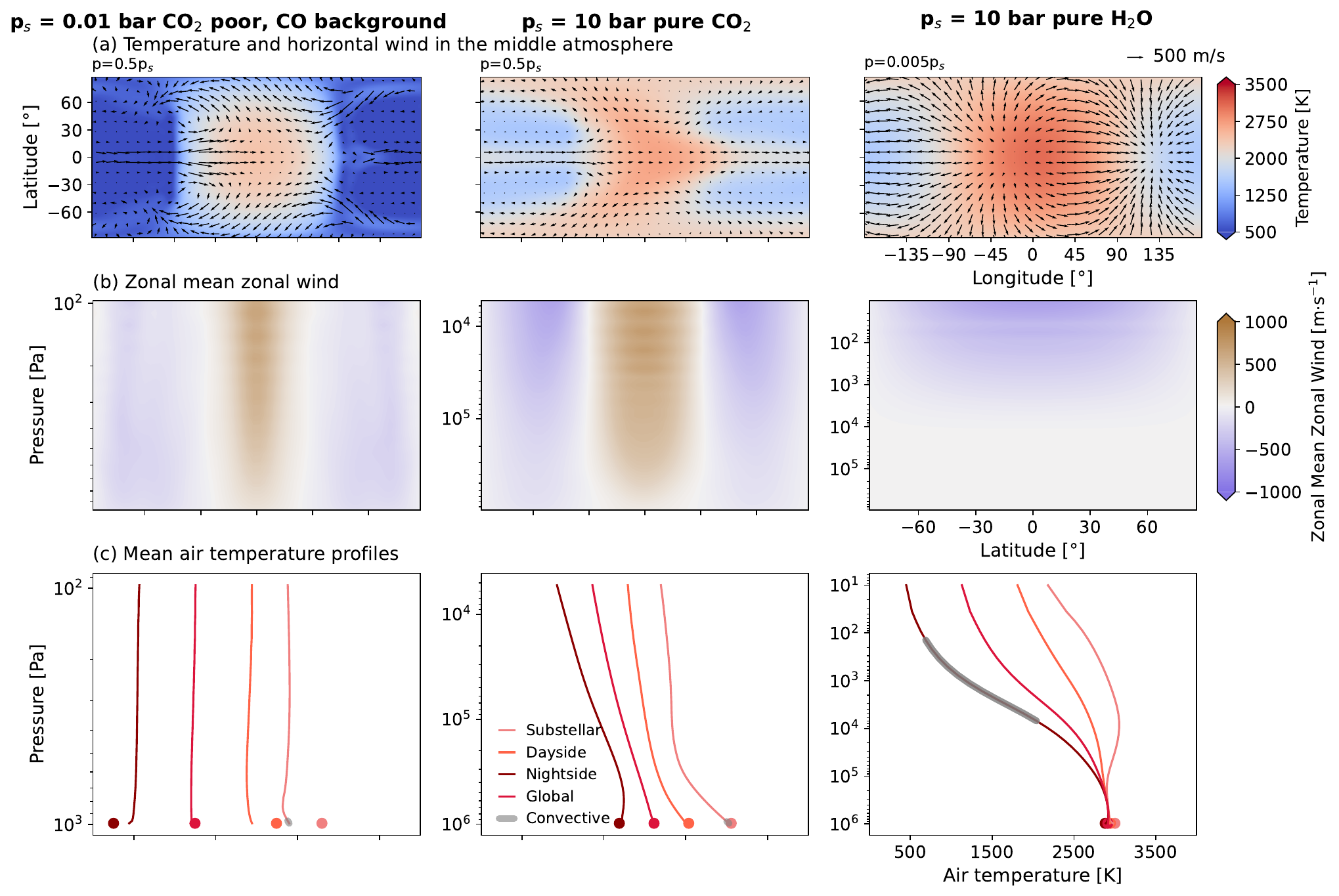}
    \caption{Long-term mean air temperature and horizontal wind (panel a), zonal mean zonal wind (panel b), and vertical temperature profiles (panel c). From left to right: a thin, CO$_2$-poor atmosphere ($p_s=0.01$ bar, CO$_2$ VMR=$10^{-8}$); a thick, pure CO$_2$ atmosphere ($p_s=10$ bar); and a thick, pure H$_2$O atmosphere ($p_s=10$ bar). (a) The thin, CO$_2$-poor and thick, pure CO$_2$ cases are plotted at the half-surface-pressure level; the pure H$_2$O case is isothermal and quiescent at depth, so the $0.005 p_s$ pressure level is plotted instead. 
    (c) Lines show different vertical air temperature profiles while colored dots at the bottom represent surface temperature. Grey shading shows convective zones, where the temperature profile matches a dry adiabat.
    For numerical stability, the thin, CO$_2$-poor simulation has a sponge layer at the top of the atmosphere (see Section \ref{sec:method}), so results in panels b and c are only shown below this pressure level.
    }
    \label{fig::atmos_dyn}
\end{figure}

Figure \ref{fig::atmos_dyn} shows the long-term mean temperature and horizontal wind fields for three representative atmospheric scenarios: a thin, CO$_2$-poor atmosphere ($p_s=0.01$ bar, CO$_2$ VMR=$10^{-8}$); a thick, pure CO$_2$ atmosphere ($p_s=10$ bar); and a thick, pure H$_2$O atmosphere ($p_s=10$ bar).

Compared with previous grey GCMs of 55 Cancri e, we find similar horizontal temperature structures but distinct vertical temperature structures. For CO$_2$- and CO-dominated atmospheres, horizontal temperatures are dominated by stationary Kelvin and Rossby waves while zonal-mean winds are marked by equatorial superrotation and mid-latitude subrotation (Figure \ref{fig::atmos_dyn}a,b). These patterns are qualitatively very similar to the grey GCMs in \citet{hammond_linking_2017}.
In contrast, the H$_2$O atmosphere exhibits a different regime: the deep atmosphere is isothermal and quiescent, while the upper atmosphere is dominated by strong convergence and divergence. This regime is likely due to H$_2$O's strong shortwave absorption. Comparing vertical temperature profiles, grey GCMs typically predict a deep convective troposphere on the dayside \citep{hammond_linking_2017}. In contrast, our non-grey simulations show that, even at the substellar point, the convective layer is vanishingly thin (Figure \ref{fig::atmos_dyn}c).

In our simulations, day-night heat redistribution is primarily governed by atmospheric thickness. This is compatible with previous theoretical predictions: a thicker atmosphere has a longer radiative timescale, which allows more efficient heat transport to the nightside \citep{showman_atmospheric_2013,koll_scaling_2022}. In contrast, molecular weight and specific heat capacity effects remain secondary across the scenarios considered here \citep{zhang_effects_2017}.

\subsection{\texorpdfstring{JWST observations favors thick and CO$_2$ rich atmospheres}{JWST observations favors thick and CO2 rich atmosphere}} \label{sec:result_spectra}
\begin{figure}[htbp]
    \centering
    \includegraphics[width=.85\textwidth]{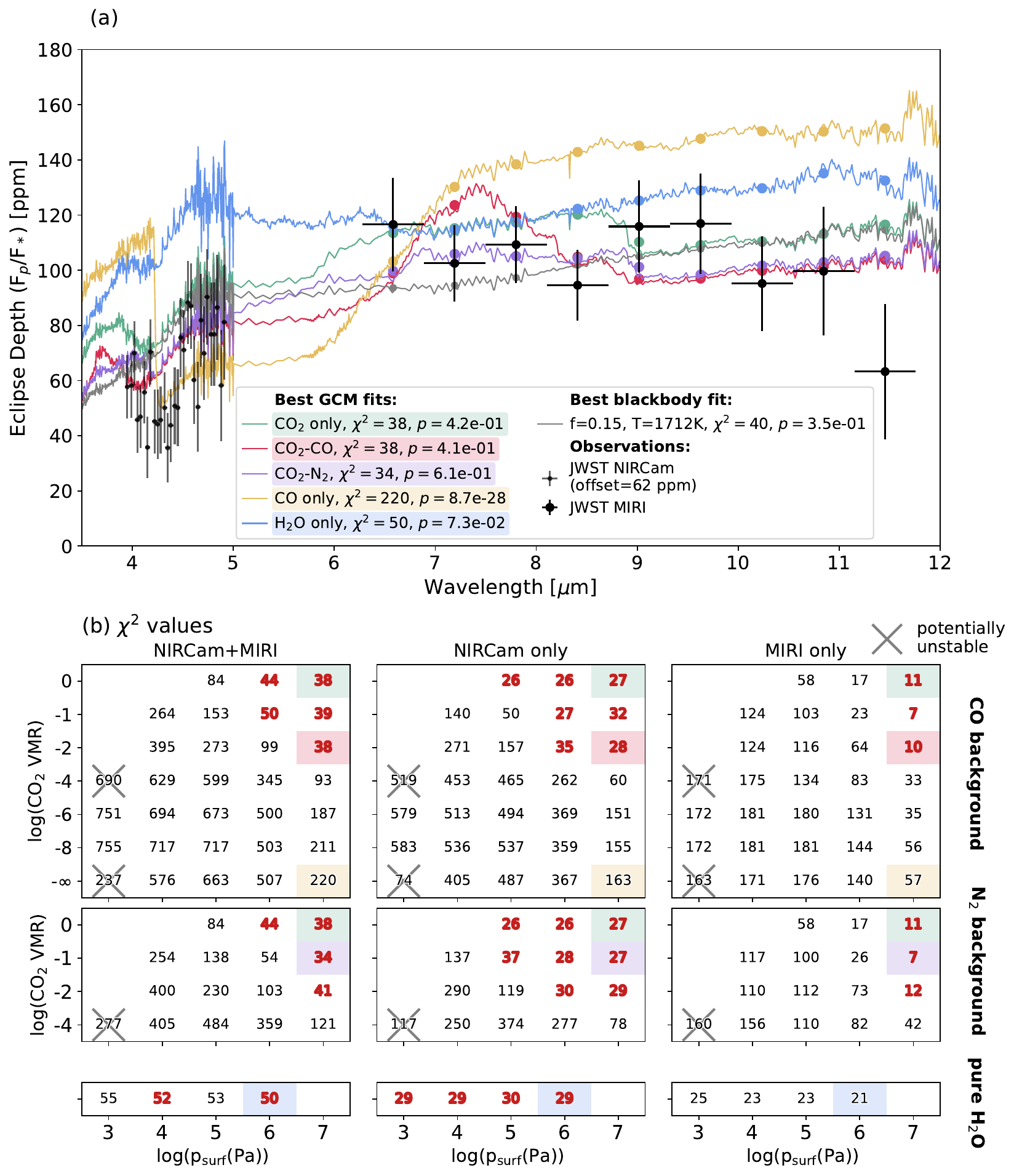}
    \caption{
    Simulated thermal emission spectra of 55 Cancri e versus JWST data from \citet{hu_secondary_2024}. The JWST data favor thick ($\ge$ 10 bar) and CO$_2$-rich ($\ge$ 10$^{-2}$ CO$_2$ by volume mixing ratio) atmospheres. 
    Panel a shows the best-fit spectra for five model categories: pure CO$_2$ (100 bar; log($p_\mathrm{surf}$(Pa)) = 7), CO$_2$-CO (100 bar, 1\% CO$_2$ VMR; log($p_\mathrm{surf}$(Pa)) = 7, log(CO$_2$ VMR) = -2), CO$_2$-N$_2$ mixed (100 bar, 10\% CO$_2$ VMR; log($p_\mathrm{surf}$(Pa)) = 7, log(CO$_2$ VMR) = -1), pure CO (100 bar; log($p_\mathrm{surf}$(Pa)) = 7), pure H$_2$O (10 bar; log($p_\mathrm{surf}$(Pa)) = 6) atmospheres, and a blackbody (f = 0.15, T = 1712 K). 
    Due to correlated noise the NIRCam vertical offset is a free parameter, for illustration NIRCam points are plotted here using an offset of 62 ppm. To facilitate direct comparison, model spectra are binned to the native resolution of the NIRCam and MIRI data (colored points).
    Panel b shows $\chi^2$ for each model based on NIRCam, MIRI, and both datasets. The colors of the shadings in panel b correspond to those of the spectra in panel a. Cases with p-value $>$ 0.05 are highlighted in red. Grey crosses show simulations that are likely unstable to atmospheric collapse (see Section \ref{sec::method_gcm}). 
    The NIRCam data favor a CO$_2$- or H$_2$O-rich atmosphere, while the MIRI data disfavor H$_2$O and strongly constrain the surface pressure to be greater than 10 bar. The combined fit to both datasets rules out several scenarios, including thin atmospheres ($<$ 10 bar) and pure-CO/CO$_2$-poor atmospheres.
    }
    
    \label{fig::eclip_spec}
\end{figure}

\begin{figure}[htbp]
    \centering
    \includegraphics[width=.55\textwidth]{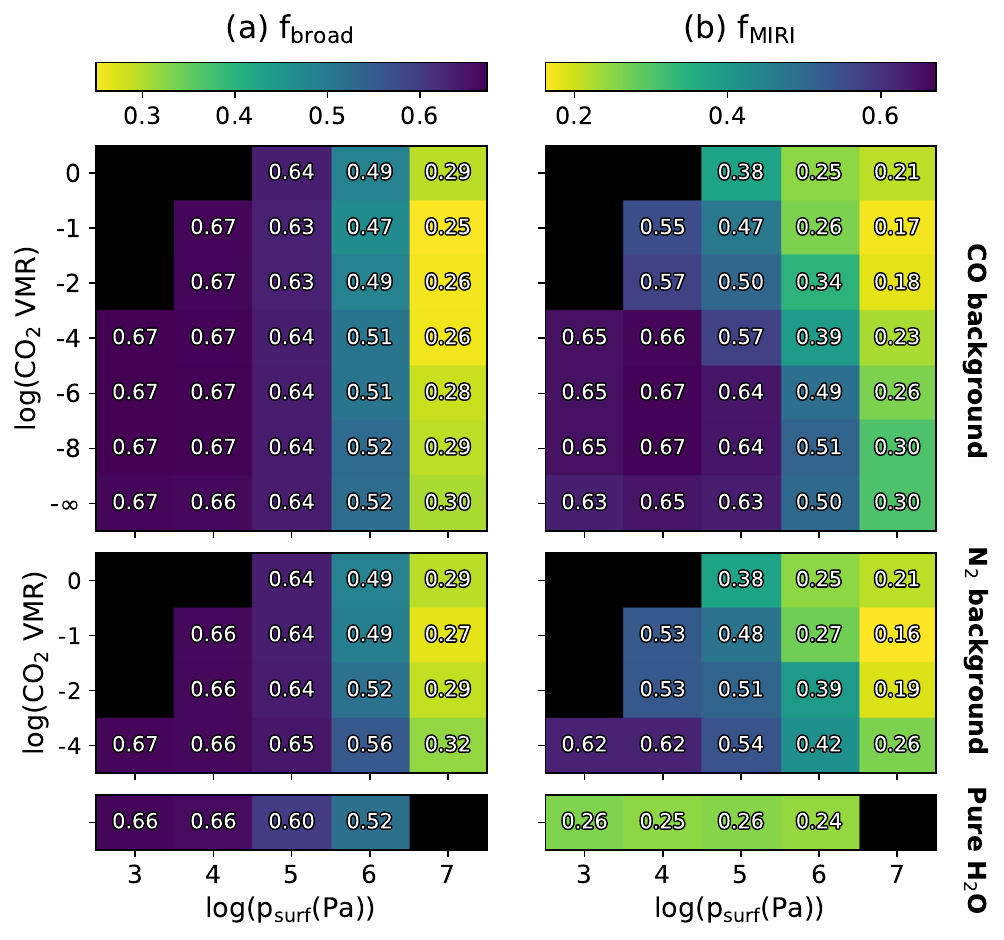}
    \caption{Broadband and MIRI band heat redistribution factors ($f$). These factors are derived via Eq. \ref{eq::brightness_temp} using brightness temperatures calculated from observer-projected fluxes $F_p$ \citep[Eq. 14]{zhan_novel_2024}. Comparing with Figure \ref{fig::eclip_spec} b, the best-fit simulations yield low MIRI-band $f$ values ($< 0.25$) due to a combination of efficient broadband heat redistribution and strong mid-infrared opacity.
    }
    \label{fig::statistic}
\end{figure}

Figure \ref{fig::eclip_spec}a compares the JWST data against our best-fit GCM simulations, which we group into 5 different categories: pure CO$_2$, CO$_2$-CO, CO$_2$-N$_2$, pure CO, and pure H$_2$O atmospheres (in colors). For reference, we also include a blackbody model (black), whose heat redistribution factor $f$ is fitted to match the data. To distinguish whether the model fits are driven by NIRCam versus MIRI, Figure \ref{fig::eclip_spec}b shows the goodness-of-fit using $\chi^2$ across our model grid for NIRCam only, MIRI only, and the combined NIRCam+MIRI dataset.

Our simulations show that the JWST data favor thick ($>10$ bar) and CO$_2$-rich atmospheres. The best fits are for CO$_2$-dominated atmospheres with 100 bar surface pressure and CO$_2$ VMRs of $1\%-100\%$ ($\chi^2 = 34$ and $38$, p-values of 0.6 and 0.4, respectively). The fits are essentially the same for thick CO$_2$-CO versus thick CO$_2$-N$_2$ atmospheres, indicating the background gas is poorly constrained.
Adopting a cutoff p-value of 0.05, CO$_2$-dominated atmospheres must have a surface pressure of at least 10 bar to match the data. Pure CO atmospheres are ruled out regardless of how thick they are. Pure H$_2$O atmospheres are borderline, with p-values right around the cutoff. However, the best fits are noticeably less good for H$_2$O than for CO$_2$-dominated atmospheres. Overall, our results thus rule out thin atmospheres ($<$ 10 bar), pure CO, and CO$_2$-poor atmospheres, which were previously proposed based on 1D models \citep{hu_secondary_2024,zilinskas_characterising_2025}.

What drives these model fits? The NIRCam data at 4-5$\mu$m primarily constrain atmospheric composition rather than heat redistribution. The observed V-shape favors CO$_2$-rich or blackbody-like (pure H$_2$O) atmospheres. While CO also absorbs near 4.5 $\mu$m, we find that CO-dominated atmospheres are ruled out because CO's opacity exhibits a sharp jump at $\sim$4.2 $\mu$m which does not match the observed V-shape (see Appendix \ref{appendix::configuration}). NIRCam provides no constraint on heat redistribution, as the absolute eclipse depth of the NIRCam data is a free parameter \citep{hu_secondary_2024}.

In contrast, MIRI data strongly constrain heat redistribution. To better understand Figure \ref{fig::eclip_spec}b, Figure \ref{fig::statistic} shows two different heat redistribution factors $f$ across our model grid, evaluated using Equation \ref{eq::brightness_temp}. First, we define $f_\mathrm{broad}$ based on the planet's broadband brightness temperature. Figure \ref{fig::statistic}a shows $f_\mathrm{broad}$ is dominated by surface pressure, with a transition from zero to full redistribution ($f_\mathrm{broad} \rightarrow 2/3$ versus $f_\mathrm{broad} \rightarrow 1/4$) between 1 and 100 bar. This is consistent with the theoretical scaling from \citet{koll_scaling_2022}. Second, defining $f_{\mathrm{MIRI}}$ based on the brightness temperature in the MIRI band only, we find that the apparent heat redistribution at a given band also depends on spectral opacity (Figure \ref{fig::statistic}b). CO$_2$ and H$_2$O absorb effectively in the MIRI band, reducing the observable brightness temperature. This allows H$_2$O- and CO$_2$-rich atmospheres to have an extremely efficient heat redistribution at MIRI wavelengths, $f_{\mathrm{MIRI}}<0.25$, even when their broadband redistribution is inefficient.

Comparing Figure \ref{fig::eclip_spec}b and \ref{fig::statistic}b shows that our best-fit GCMs all have efficient heat redistribution at MIRI wavelengths ($f_{\mathrm{MIRI}} < 0.25$). This means the MIRI data require the planet to be colder than a zero-albedo object with full heat redistribution, consistent with the best-fit heat redistribution factor we find for a blackbody model ($f=0.15$). Assuming a low planetary albedo, such a low brightness temperature requires a combination of efficient broadband heat redistribution and significant mid-infrared opacity. Consequently, the data are best fitted by thick ($\ge$ 10 bar) and CO$_2$-rich ($>1\%$ CO$_2$) atmospheres.

\subsection{Time Variability Due to Large-scale Atmospheric Dynamics is Much Weaker Than Observed}
\label{sec:result_variability}
\begin{figure}[htbp]
    \centering
    \includegraphics[width=.92\textwidth]{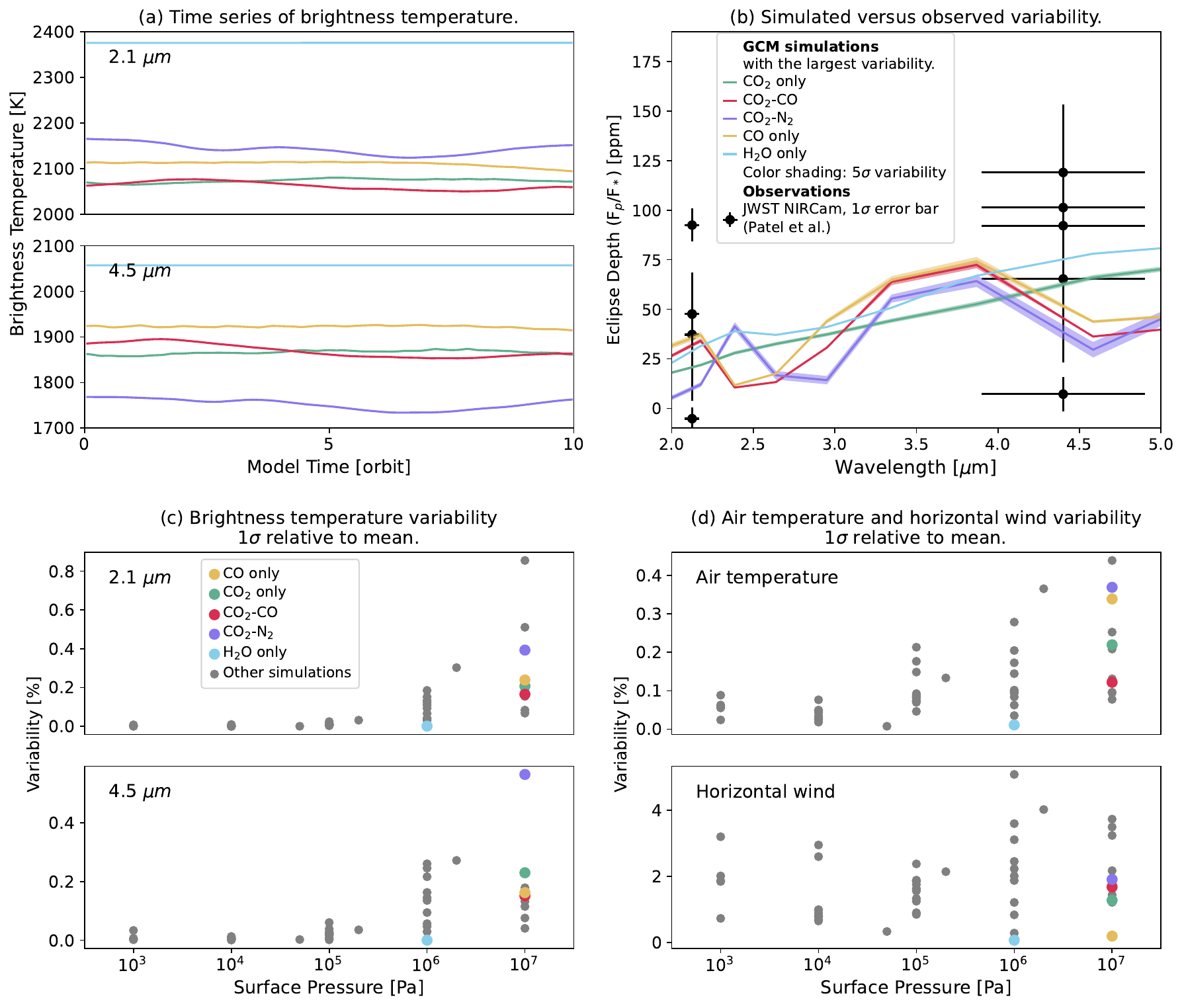}
    \caption{Time variability due to large-scale atmospheric dynamics is much weaker than observations \citep{patel_jwst_2024}. Panel a shows the time series of brightness temperature at 2.1 $\mu$m and 4.5 $\mu$m. Panel b shows the variability of eclipse spectra and the comparison with observations. Panel a and b shows the simulations with the largest variability in the 4.5 $\mu$m bandpass in each scenario. Note that this selection differs from Figure \ref{fig::eclip_spec}, which displays the simulations that best fit the JWST emission spectra. 
    Panel c shows the brightness temperature at 2.1 $\mu$m and 4.5 $\mu$m changes with surface pressure of each simulations. Panel d shows the global mean air temperature and horizontal wind at half-surface-pressure level. Both panel c and d are 1 $\sigma$ standard deviation relative to mean. Colored lines shows simulations with largest brightness temperature in each scenario as labeled in panel b. Lines and points with the same color in the four panels represent the same simulation. We analyze the time variability of 10 model orbits after the simulations reach statistical equilibrium.  }
    \label{fig::variability}
\end{figure}

Our simulations suggest that any variability induced by large-scale atmospheric dynamics is significantly weaker than that observed. To analyze time-variability we group all GCMs again into 5 categories: pure CO$_2$, CO$_2$-CO, CO$_2$-N$_2$, pure CO, and pure H$_2$O. In each category, we then pick the simulation which has the largest variability at 4.5 $\mu$m. Figure \ref{fig::variability}a shows time series of brightness temperatures of these simulations, while Figure \ref{fig::variability}b compares their simulated eclipse spectra against the JWST data \citep[][Figure 10]{patel_jwst_2024}. Here, colored shaded regions and black error bars show the variability in eclipse depths for the GCMs versus observations. We find that the simulated time series are nearly flat, and the eclipse depth variability in the GCMs is much narrower than in the observations.

Despite the fact that none of our GCM simulations produce anything like the observed variability, a more thorough analysis shows that variability is generally stronger in thicker atmospheres. Figure \ref{fig::variability}c shows a scatter plot of the relative variability of brightness temperature as a function of surface pressure. There is a clear trend in which brightness temperature becomes more variable in thicker atmospheres.

Brightness temperature becomes more variable because both static and kinetic energy can vary more in thicker atmospheres.
We evaluate static and kinetic energy by root-mean-square (rms) of air temperature $T(t)$ and horizontal wind velocity $U(t)$ at the half-surface-pressure level, defined as 
$T(t)=\sqrt{\int T^2(\lambda,\varphi,t)\frac{dA(\lambda,\varphi)}{A}},$
and 
$U(t)=\sqrt{\int (u^2(\lambda,\varphi,t)+v^2(\lambda,\varphi,t))\frac{dA(\lambda,\varphi)}{A}}.$
Here, $T(\lambda,\varphi,t)$, $u(\lambda,\varphi,t)$, $u(\lambda,\varphi,t)$ are time series of air temperature, zonal wind and meridional wind at longitude index $\lambda$, latitude index $\varphi$ at the half-surface-pressure level. We then calculate the standard deviations relative to mean of the time series $T(t)$ and $U(t)$, e.g., $\sigma_T = \frac{1}{\bar{T}}\sqrt{ \frac{1}{t_\mathrm{end} - t_\mathrm{start}} \int_{t_\mathrm{start}}^{t_\mathrm{end}} \left( T(t) - \bar{T} \right)^2 \, dt } $. Figure \ref{fig::variability}d shows the 1$\sigma$ variability as a function of surface pressure of our simulations. We find the variability of $T$, which represents static energy, and $U$, which represents kinetic energy, generally increase with surface pressure. This is because the radiative timescale $\tau_{\mathrm{rad}}=\frac{c_pp_s}{4\sigma T_e^3g}$, increases with the surface pressure $p_s$. 55 Cancri e is extremely hot, so waves are generally quickly damped out by radiative cooling ($\tau_\mathrm{wave} \gg \tau_\mathrm{rad}$). However, the ratio $\tau_\mathrm{wave}/\tau_\mathrm{rad}$ decreases with surface pressure, from $\sim 10^3$ for our thinnest atmospheres to $\sim 0.1$ for the thickest ones.

\section{Discussion}
\label{sec:discussion}
Our analysis provides compelling evidence that the atmosphere of 55 Cancri e is significantly thicker and richer in $\text{CO}_2$ than previously proposed. The fact that such a substantial atmosphere is able to persist against atmospheric escape suggests 55 Cancri e formed with far more carbon than the Solar System rocky planets. Assuming energy-limited escape and a relatively high escape efficiency, 55 Cancri e should have lost about $3\%$ of its mass in volatiles over its lifetime \citep{hu_secondary_2024}. Alternatively, based on CO$_2$ hydrodynamic escape models, 55 Cancri should have lost about $0.5\%$ of its mass in carbon \citep{tian_thermal_2009,ji_cosmic_2025}. These numbers should be compared to the bulk carbon contents of rocky planets in the Solar System. Earth contains only 0.01–0.1 wt\% carbon \citep{fischer_carbon_2020}. Venus's thick CO$_2$-rich atmosphere suggests a similar observable carbon inventory as Earth's, though the carbon content of Venus' interior is still unknown \citep{lecuyer_comparison_2000,halliday_origins_2013,avice_noble_2022}. Assuming 55 Cancri e's inferred thick CO$_2$ atmosphere persisted over geologic timescales, the planet would need to have formed with at least 5--30 times more carbon than Earth and Venus to still maintain such an atmosphere today. These numbers are a lower bound, because a thick carbon-rich atmosphere should be equilibrated with even more carbon in the planet's surface lava ocean and interior.

Our simulations also show that clear-sky atmospheric dynamics alone are insufficient to reproduce the strong observed time variability. A thick atmosphere disfavors the transient outgassing mechanism previously proposed by \citet{heng_transient_2023}. The atmospheric escape timescale for a $> 10$ bar atmosphere far exceeds the short variability timescales observed by JWST \citep{patel_jwst_2024}. Our results thus favor alternative variability mechanisms, such as magma-cloud feedbacks \citep{loftus_extreme_2024}, tidal heating \citep{farhat_magma_2026},
episodic volcanic dust injection \citep{meier_interior_2023}, or a circumstellar torus \citep{demory_variability_2016}.

\section{Conclusion}
\label{sec:conclusion}
We simulate the atmosphere of 55 Cancri e using a 3D non-grey GCM with custom high-temperature correlated-$k$ coefficients and reinterpret previous JWST observations. Our key findings are:

\begin{itemize}
    \item \textbf{Atmospheric Dynamics:} Our non-grey GCM simulations reveal significantly different vertical profiles compared to grey GCMs. CO$_2$-rich GCMs exhibit stationary Rossby and Kelvin waves similar to grey GCMs, while H$_2$O GCMs feature upper-level divergence and convergence overlying a deep quiescent layer.
    
    \item \textbf{Atmospheric Characterization:} Current JWST data favor thick ($\ge 10$ bar) and CO$_2$-rich ($> 1\%$ VMR) atmospheres, driven by efficient heat redistribution and strong gas absorption in the MIRI bandpass. We rule out thin or CO$_2$-poor scenarios proposed based on 1D models.
    
    \item \textbf{Eclipse Variability:} Large-scale atmospheric dynamics cannot explain the observed variability. Although simulated variability increases with surface pressure, even 100 bar atmospheres show far weaker variability than observed. It also takes too long for a thick atmosphere to escape to space, which suggests the planet's variability is unlikely caused by transient outgassing \citep{heng_transient_2023}, favoring alternative mechanisms (e.g., clouds).
\end{itemize}

Upcoming eclipse and transmission spectroscopy observations with JWST MIRI/MRS and NIRISS/SOSS (GO programs 7875, 9825, and 12237).

are scheduled to probe 55 Cancri e's atmosphere, which should constrain our conclusions about the atmosphere's thickness and composition even more precisely.

\section{Acknowledgments}

We thank Stephen I. Thomson for developing Isca and making it publicly available, the UK Met Office for developing SOCRATES, and Eric Wolf for helpful advice on correlated-k methods. We thank the authors of \citet{hu_secondary_2024}, \citet{zilinskas_characterising_2025}, and \citet{patel_jwst_2024} for making their processed data publicly available, which were originally derived from JWST observations hosted at the Mikulski Archive for Space Telescopes (MAST). We also thank Yichen Gao, Yueyun Ouyang, Haolin Li, Jun Yang, Feng Ding, and Xuan Ji for insightful discussions. D.D.B.K. acknowledges support from the National Natural Science Foundation of China (NSFC) under grant number 42250410318. The scripts used to generate our correlated-k tables are available at https://github.com/ruizhizhan/SocSpecGen, where the latest version is maintained; the version associated with this article is archived at https://doi.org/10.5281/zenodo.20369151. All data necessary to reproduce this study, including the absorption cross sections, SOCRATES spectral files, simulation outputs, and the analysis scripts used to generate the figures, are publicly available at https://doi.org/10.5281/zenodo.19388488.

\clearpage
\appendix

\setcounter{figure}{0}
\setcounter{table}{0}

\renewcommand{\thefigure}{\thesection\arabic{figure}}
\renewcommand{\thetable}{\thesection\arabic{table}}

\section{Non-grey Radiative Transfer}

\label{appendix::configuration}
We calculate absorption cross-sections from ExoMol line lists at $0.01\,\text{cm}^{-1}$ resolution with Voigt profiles using \texttt{ExoCross} \citep{yurchenko_exocross_2018}, applying air- or self-broadening where available. Gas species include CO$_2$ \citep{rothman_energy_1992,yurchenko_exomol_2022}, H$_2$O \citep{polyansky_exomol_2018}, CO including quadrupole transitions \citep{li_rovibrational_2015,somogyi_calculation_2021,faure_pressure_2013,gordon_hitran2016_2017,guest_predicting_2024}, and N$_2$ \citep{western_spectrum_2018,western_pgopher_2017,shemansky_n2_1969,jans_rovibronic_2024,gamache_thermodynamic_2023}. For each gas, we only consider the dominant isotopologue. The ExoMol opacities are supplemented with UV absorption data \citep{souza_photoabsorption_1994,keller-rudek_mpi-mainz_2013} (Fateev et al., in press) and CIA data for CO$_2$--CO$_2$ and N$_2$--N$_2$ \citep{karman_update_2019}. 

This appendix first shows representative gas absorption cross sections across a wide range of pressure and temperature (Section \ref{appendix::cross_section}). We then validate our non-grey radiative transfer by testing and optimizing the configurations of our correlated-$k$ tables (Section \ref{appendix::configuration}), and comparing our radiative calculations against line-by-line calculations (Section \ref{appendix::lbl}). Finally, for atmospheres with multiple gases, we also validate the assumption used to treat opacity overlap (Section \ref{appendix::gas_overlap}).

\subsection{Absorption cross sections}
\label{appendix::cross_section}

\begin{figure}[h!]
    \centering
    \includegraphics[width=.85\textwidth]{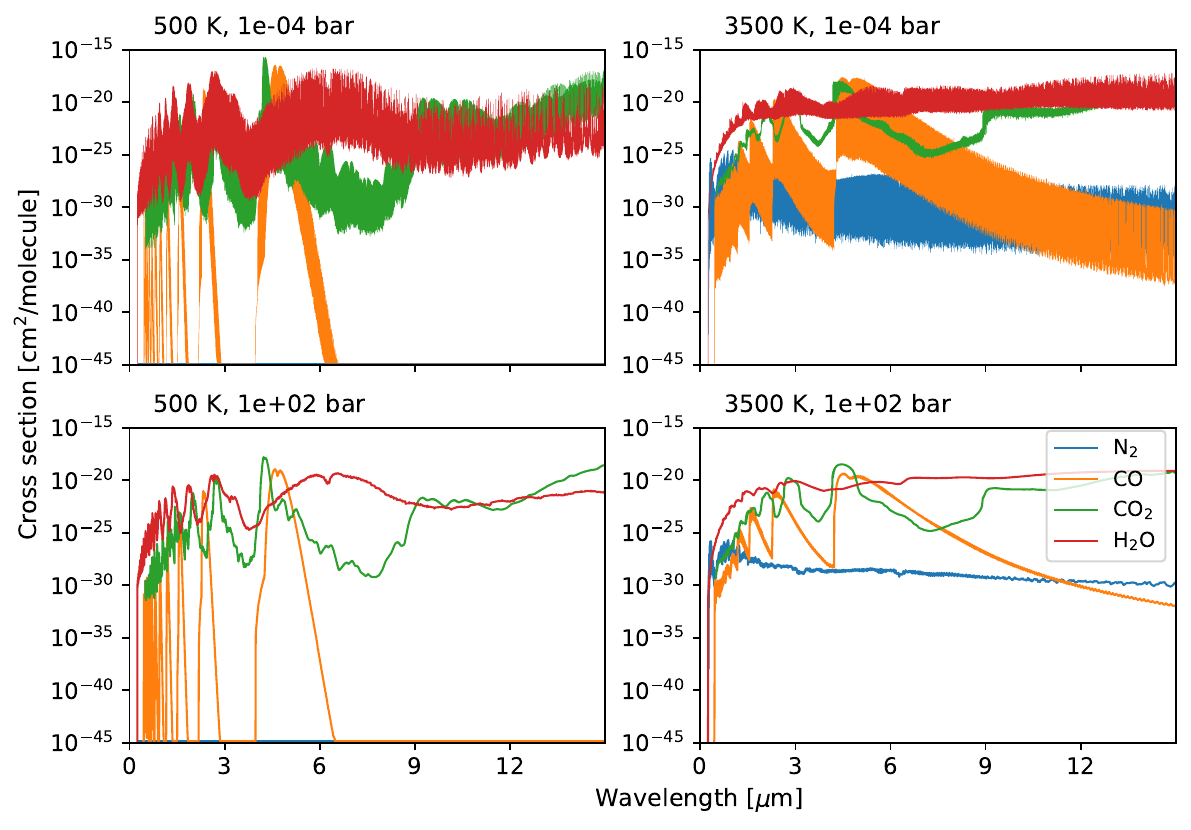}
    \caption{Absorption cross sections of N$_2$, CO, CO$_2$ and H$_2$O at 500 K and 3500 K, 10 Pa and 100 bar. These cross sections are used to generate correlated-$k$ coefficients used in this work.} 
    \label{fig::xsec_appen}
\end{figure}
  
Figure \ref{fig::xsec_appen} shows representative absorption cross-sections for N$_2$, CO, CO$_2$, and H$_2$O, computed using \texttt{ExoCross}. As expected, these calculations closely resemble previous ExoMol opacity calculations \citep{tennyson_exomol_2018}.

\subsection{Configurations of correlated-$k$ Tables}

\begin{figure}[h!]
    \centering
    \includegraphics[width=.98\textwidth]{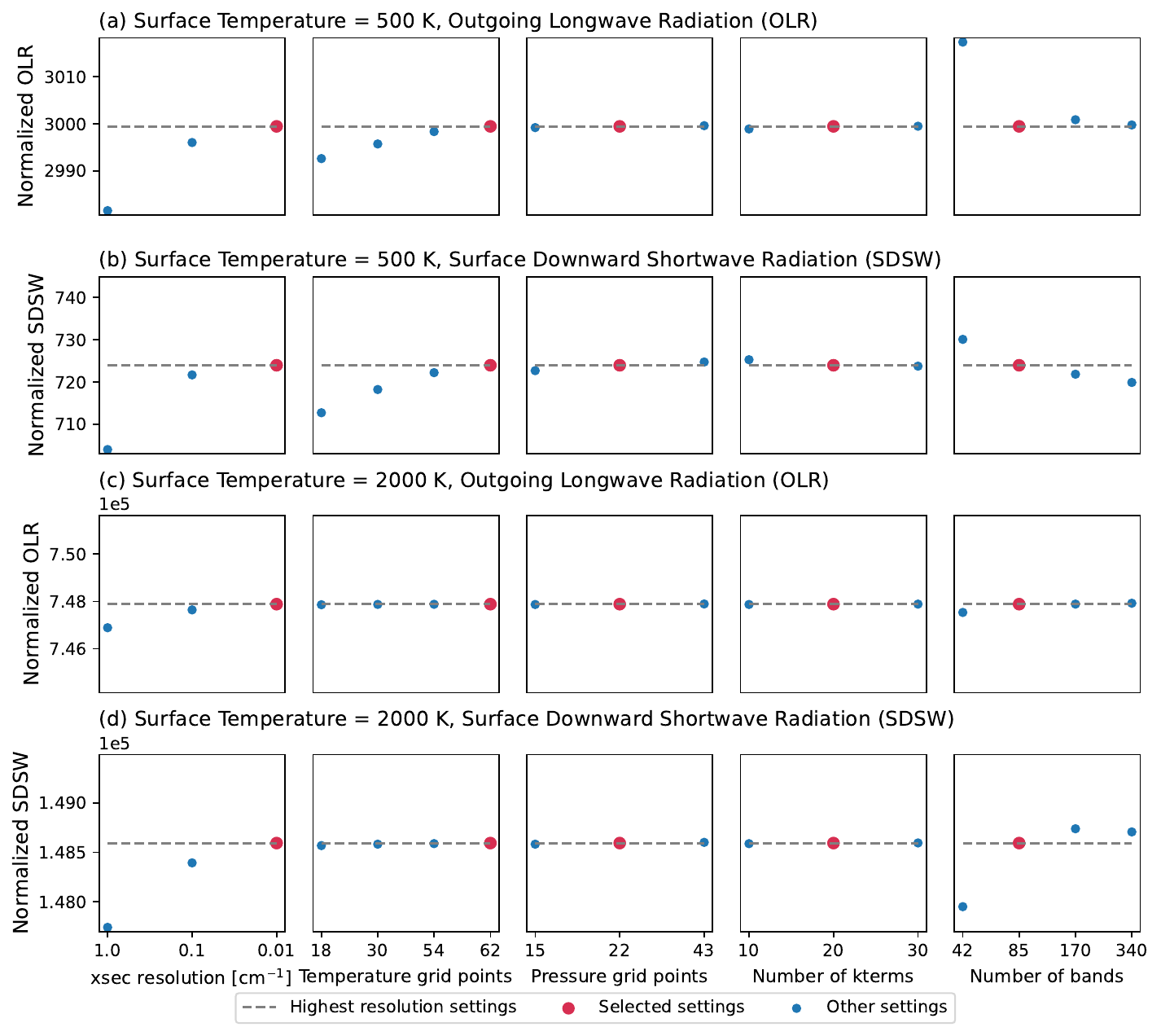}
    \caption{Validations for configurations of correlated-$k$ tables.} 
    \label{fig::ck_settings}
\end{figure}

Next, we test different correlated-$k$ table configurations to balance accuracy and computational efficiency. The tests use \texttt{SOCRATES} with a pure CO$_2$ atmosphere and a fixed 1D temperature-pressure (T-p) profile, and compare the convergence of outgoing longwave radiation (OLR) and surface downward shortwave radiation (SDSW). To cover a large range of temperatures, we test two surface temperatures (500 K and 2000 K). In each case, the atmospheric T-p profile is set to a dry adiabat in the lower atmosphere and an isothermal layer higher up.
We test convergence with respect to the following parameters: pre-computed cross section (xsec) resolution, number of grid points in the temperature, number of grid points in the pressure grid, maximum number of k-terms, and number of spectral bands. 

Figure \ref{fig::ck_settings} shows the results. We find that OLR and SDSW are most sensitive to xsec resolution and the number of spectral bands. At lower temperatures, the number of temperature grid points also matters. Interestingly, OLR and SDSW are quite insensitive to the number of pressure grid points and the number of k-terms (third and fourth columns in Figure \ref{fig::ck_settings}). Both quantities vary by less than 1\% between the computationally most expensive and least expensive setting.

The sensitivities in Figure \ref{fig::ck_settings} have to be balanced against the numerical cost of increasing a given parameter. Generally, the larger a parameter is, the more accurate the resulting radiative transfer calculation.
\begin{itemize}
\item Xsec resolution doesn't affect the GCM efficiency at all, so we choose the highest resolution here.
\item For the temperature grid, we choose the highest resolution. The grid points are distributed as follows: intervals of 25 K between 100 K and 500 K, 50 K between 500 K and 1000 K, and 100 K between 1000 K and 4500 K.
\item For the pressure grid, moderate resolution is sufficient. The pressure grid points are logarithmically distributed between 1 Pa and 100 bar.
\item For the number of kterms, moderate resolution is again sufficient.
\item For the number of spectral bands, doubling the number of bands generally slows down our GCM simulations by a factor of two. Based on Figure \ref{fig::ck_settings}, we choose 85 bands, as adding even more bands improves accuracy only by a small ($<1\%$) amount. The bands are distributed as follows: a uniform grid of 400 cm$^{-1}$ between 0 and 20000 cm$^{-1}$, followed by a 2000 cm$^{-1}$ grid from 20000 up to 90000 cm$^{-1}$. Appropriate for the extremely high temperatures on 55 Cancri e, we use the same band distribution for both shortwave and longwave calculations.
\end{itemize}

Based on these tests, our final k-table configuration is: 0.01 cm$^{-1}$ for the xsec resolution, 62 points for the temperature grid, 22 points for the pressure grid, 20 for the maximum number of k-terms, and 85 spectral bands.

\subsection{Validation Against Line-by-Line Calculations}
\label{appendix::lbl}

\begin{figure}[h!]
    \centering
    \includegraphics[width=.7\textwidth]{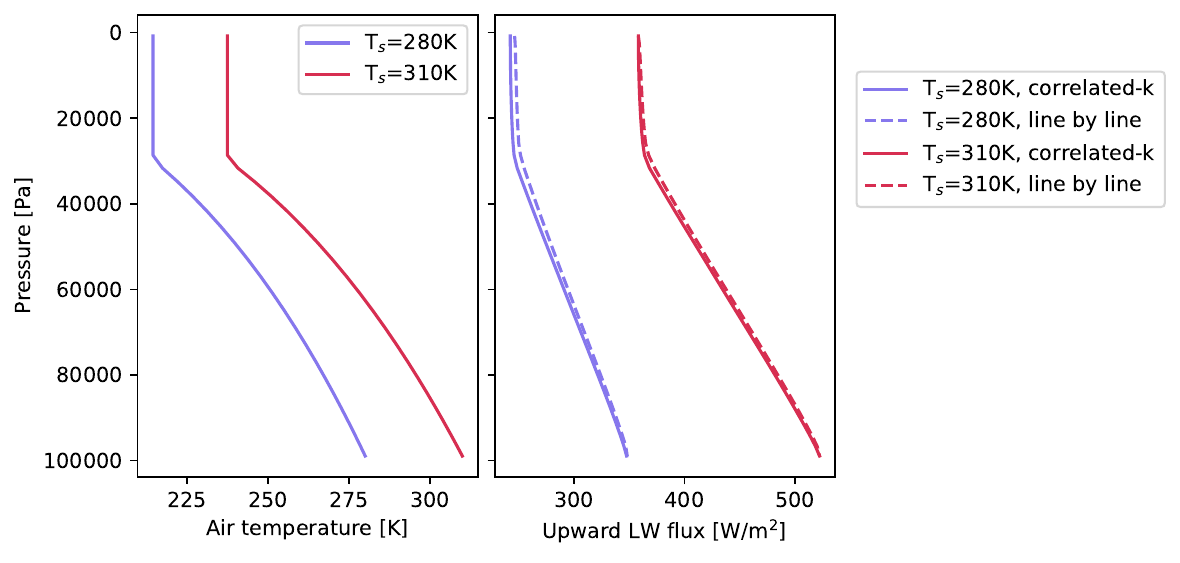}
    \caption{Temperature profiles and upward longwave flux using correlated-$k$ methond and line-by-line calculations. On the left panel, the red and purple lines represent air temperature of 280 K and 310 K surface temperature scenario, respectively. On the right panel, the solid line and dashed line represent correlated-$k$ and line-by-line calculations. The comparison suggests that the difference between correlated-$k$ and line-by-line is negligible.} 
    \label{fig::lbl_validation}
\end{figure}

Next we validate our non-grey radiative transfer and custom correlated-$k$ coefficients against a line-by-line radiative transfer model, \texttt{PyRADS} \footnote{PyRADS on GitHub: https://github.com/danielkoll/PyRADS}. \texttt{PyRADS} is based on HITRAN2016 line lists. Since ExoMol contains many orders of magnitude more absorption lines than HITRAN2016, which generally become more important at higher temperature, we therefore limit the comparison to an Earth-like temperature regime. For simplicitiy, collision induced absorption is also excluded. We fix the vertical T-p profile and compare the resulting upward longwave flux profile. As shown in Figure \ref{fig::lbl_validation}, the two radiative transfer codes agree with each other to within a few percent.

\subsection{Gas Overlap Assumption}
\label{appendix::gas_overlap}

\begin{figure}[h!]
    \centering
    \includegraphics[width=.4\textwidth]{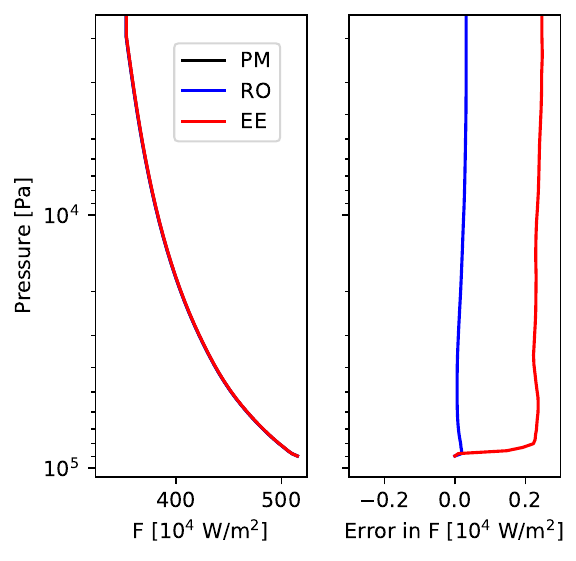}
    \caption{Upward longwave flux calculated under different overlap assumptions for an atmospheric composition of 50\% CO$_2$ and 50\% CO. Here, 'PM' denotes pre-mixed cross sections, 'RO' stands for random overlap, and 'EE' refers to equivalent extinction, which is the method adopted in our simulations. The three curves in the left panel are nearly identical, while the right panel shows the differences between 'EE' and 'RO' relative to 'PM'. 'RO' is slightly more accurate than 'EE', but 'EE' is sufficiently accurate and more efficient.}
    \label{fig::overlap_assumption_flux}
\end{figure}

\begin{table}[h!]
    \centering
    \setlength\tabcolsep{12pt}
    \begin{tabular}{cccc}
    \hline\hline
     & Pre-mixed & Random Overlap & Equivalent Extinction\\\hline
     Global Mean OLR [W/m$^2$] & 880074.24 & 870178.12 & 877607.24\\
     \hline
    \end{tabular}
    \caption{Global mean outgoing longwave radiation (OLR) for to the atmospheres shown in Figure \ref{fig::overlap_assumption_flux}.}
    \label{tab:overlap}
\end{table}

For atmospheres with mixed-gas compositions, previous work tended to use the equivalent extinction method \citep[e.g.][]{turbet_trappist-1_2022}.
To make sure our results are robust to this assumption, we compare the equivalent extinction method with two alternatives: random overlap and pre-mixed. For the most accurate pre-mixed method, we follow \citet{amundsen_treatment_2017} and compute the total absorption coefficient $\kappa^{tot}(\nu,p,T)$ by summing the line-by-line absorption coefficients for all absorbing species weighted by their relative abundances,
\begin{equation}
    \kappa^{tot}(\nu,p,T)=\sum_{i=1}^{N_s}\zeta_i(p,T)\kappa_i(\nu,p,T),
\end{equation}
where $\zeta_i(p,T)$ and $\kappa_i(\nu,p,T)$ are the mixing ratio of gas $i$ at pressure $p$ and temperature $T$ and absorption coefficient at frequency $\nu$ respectively. In our simulations gases are assumed well-mixed, so $\zeta_i(p,T)$ is a constant. We then generate correlated-$k$ coefficients based on the mixed absorption coefficients and compute upward longwave fluxes as well as OLR.

As shown in Figure \ref{fig::overlap_assumption_flux} and Table \ref{tab:overlap}, the difference between different overlap assumptions affects vertical flux profiles and OLR by less than 1\%. Our results agree with \citet{amundsen_treatment_2017}, who found that the random overlap method is slightly more accurate than equivalent extinction. However, as equivalent extinction is already good enough and much more efficient, we choose this method for our simulations.

\clearpage

\section{Sensitivity to Horizontal Resolution}
\label{appendix::horizontal_res}

\begin{figure}[h!]
    \centering
    \includegraphics[width=1.0\textwidth]{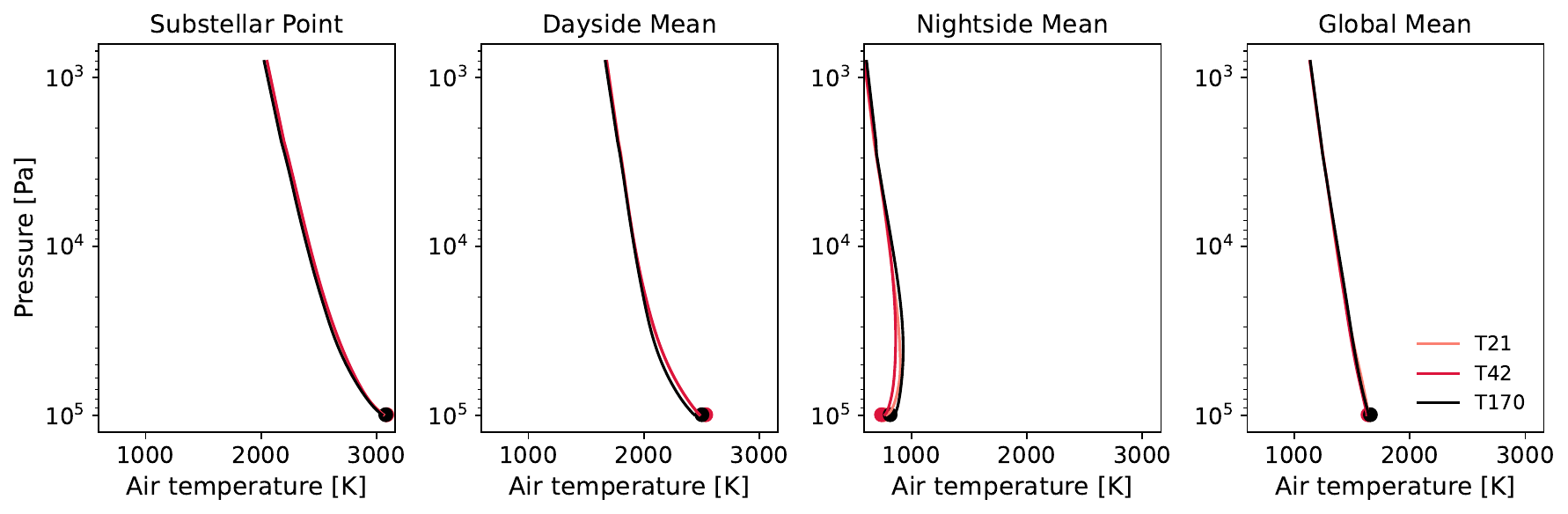}
    \caption{Vertical temperature profiles of substellar point, dayside mean, nightside mean and global mean of simulations with horizontal resolution `T21' ($5.625^\circ \times 5.625^\circ$), `T42' ($2.8125^\circ \times 2.8125^\circ$), and `T170' ($0.7031^\circ \times 0.7031^\circ$). The difference between low and high resolutions is negligible. We adopt the moderate `T42' resolution for most simulations, while using the lower `T21' resolution only for the computationally more demanding pure H$_2$O cases.}
    \label{fig::horizontal_res}
\end{figure}

To ensure our simulations use a sufficiently high resolution, we conducted a series of sensitivity tests across three horizontal resolutions: `T21' ($5.625^\circ \times 5.625^\circ$), `T42' ($2.8125^\circ \times 2.8125^\circ$), and `T170' ($0.7031^\circ \times 0.7031^\circ$). As illustrated in Figure \ref{fig::horizontal_res}, the vertical temperature profiles at the substellar point, as well as the dayside, nightside, and global means, show negligible variations across the different grids. Given this robust consistency, we adopt the moderate `T42' resolution for the majority of our simulations to balance numerical accuracy with computational cost. For pure H$_2$O atmospheres, which are computationally more demanding, we utilize the `T21' resolution.

\bibliography{article}

@article{penn_atmospheric_2018,
	title = {Atmospheric {Circulation} and {Thermal} {Phase}-curve {Offset} of {Tidally} and {Nontidally} {Locked} {Terrestrial} {Exoplanets}},
	volume = {868},
	issn = {1538-4357},
	url = {https://iopscience.iop.org/article/10.3847/1538-4357/aaeb20},
	doi = {10.3847/1538-4357/aaeb20},
	language = {en},
	number = {2},
	urldate = {2022-09-06},
	journal = {The Astrophysical Journal},
	author = {Penn, James and Vallis, Geoffrey K.},
	month = dec,
	year = {2018},
	pages = {147},
}

@article{zhang_effects_2017,
	title = {Effects of {Bulk} {Composition} on {The} {Atmospheric} {Dynamics} on {Close}-in {Exoplanets}},
	volume = {836},
	issn = {1538-4357},
	url = {http://arxiv.org/abs/1607.04260},
	doi = {10.3847/1538-4357/836/1/73},
	language = {en-US},
	number = {1},
	urldate = {2023-02-23},
	journal = {The Astrophysical Journal},
	author = {Zhang, Xi and Showman, Adam P.},
	month = feb,
	year = {2017},
	note = {arXiv:1607.04260 [astro-ph]},
	pages = {73},
}

@article{koll_deciphering_2015,
	title = {{DECIPHERING} {THERMAL} {PHASE} {CURVES} {OF} {DRY}, {TIDALLY} {LOCKED} {TERRESTRIAL} {PLANETS}},
	volume = {802},
	issn = {0004-637X},
	url = {https://dx.doi.org/10.1088/0004-637X/802/1/21},
	doi = {10.1088/0004-637X/802/1/21},
	language = {en},
	number = {1},
	urldate = {2023-07-12},
	journal = {The Astrophysical Journal},
	publisher = {The American Astronomical Society},
	author = {Koll, Daniel D. B. and Abbot, Dorian S.},
	month = mar,
	year = {2015},
	pages = {21},
}

@article{koll_scaling_2022,
	title = {A {Scaling} for {Atmospheric} {Heat} {Redistribution} on {Tidally}-{Locked} {Rocky} {Planets}},
	volume = {924},
	issn = {0004-637X, 1538-4357},
	url = {http://arxiv.org/abs/1907.13145},
	doi = {10.3847/1538-4357/ac3b48},
	number = {2},
	urldate = {2023-10-25},
	journal = {The Astrophysical Journal},
	author = {Koll, Daniel D. B.},
	month = jan,
	year = {2022},
	note = {arXiv:1907.13145 [astro-ph]},
	pages = {134},
}

@article{turbet_trappist-1_2022,
	title = {The {TRAPPIST}-1 {Habitable} {Atmosphere} {Intercomparison} ({THAI}). {Part} {I}: {Dry} {Cases} -- {The} fellowship of the {GCMs}},
	volume = {3},
	issn = {2632-3338},
	shorttitle = {The {TRAPPIST}-1 {Habitable} {Atmosphere} {Intercomparison} ({THAI}). {Part} {I}},
	url = {http://arxiv.org/abs/2109.11457},
	doi = {10.3847/PSJ/ac6cf0},
	number = {9},
	urldate = {2024-01-10},
	journal = {The Planetary Science Journal},
	author = {Turbet, Martin and Fauchez, Thomas J. and Sergeev, Denis E. and Boutle, Ian A. and Tsigaridis, Kostas and Way, Michael J. and Wolf, Eric T. and Domagal-Goldman, Shawn D. and Forget, François and Haqq-Misra, Jacob and Kopparapu, Ravi K. and Lambert, F. Hugo and Manners, James and Mayne, Nathan J. and Sohl, Linda},
	month = sep,
	year = {2022},
	note = {arXiv:2109.11457 [astro-ph, physics:physics]},
	pages = {211},
}

@article{fu_correlated_1992,
	chapter = {Journal of the Atmospheric Sciences},
	title = {On the {Correlated} k-{Distribution} {Method} for {Radiative} {Transfer} in {Nonhomogeneous} {Atmospheres}},
	volume = {49},
	issn = {0022-4928, 1520-0469},
	url = {https://journals.ametsoc.org/view/journals/atsc/49/22/1520-0469_1992_049_2139_otcdmf_2_0_co_2.xml},
	doi = {10.1175/1520-0469(1992)049<2139:OTCDMF>2.0.CO;2},
	language = {EN},
	number = {22},
	urldate = {2024-03-11},
	journal = {Journal of the Atmospheric Sciences},
	publisher = {American Meteorological Society},
	author = {Fu, Qiang and Liou, K. N.},
	month = nov,
	year = {1992},
	pages = {2139--2156},
}

@article{selsis_thermal_2011,
	title = {Thermal phase curves of nontransiting terrestrial exoplanets. {I}. {Characterizing} atmospheres},
	volume = {532},
	issn = {0004-6361},
	url = {https://ui.adsabs.harvard.edu/abs/2011A&A...532A...1S},
	doi = {10.1051/0004-6361/201116654},
	urldate = {2024-03-26},
	journal = {Astronomy and Astrophysics},
	author = {Selsis, F. and Wordsworth, R. D. and Forget, F.},
	month = aug,
	year = {2011},
	note = {ADS Bibcode: 2011A\&A...532A...1S},
	pages = {A1},
}

@article{showman_atmospheric_2013,
	title = {{ATMOSPHERIC} {DYNAMICS} {OF} {BROWN} {DWARFS} {AND} {DIRECTLY} {IMAGED} {GIANT} {PLANETS}},
	volume = {776},
	issn = {0004-637X},
	url = {https://dx.doi.org/10.1088/0004-637X/776/2/85},
	doi = {10.1088/0004-637X/776/2/85},
	language = {en},
	number = {2},
	urldate = {2024-04-08},
	journal = {The Astrophysical Journal},
	publisher = {The American Astronomical Society},
	author = {Showman, Adam P. and Kaspi, Yohai},
	month = oct,
	year = {2013},
	pages = {85},
}

@article{amundsen_treatment_2017,
	title = {Treatment of overlapping gaseous absorption with the correlated- \textit{k} method in hot {Jupiter} and brown dwarf atmosphere models},
	volume = {598},
	issn = {0004-6361, 1432-0746},
	url = {http://www.aanda.org/10.1051/0004-6361/201629322},
	doi = {10.1051/0004-6361/201629322},
	language = {en},
	urldate = {2024-05-08},
	journal = {Astronomy \& Astrophysics},
	author = {Amundsen, David S. and Tremblin, Pascal and Manners, James and Baraffe, Isabelle and Mayne, Nathan J.},
	month = feb,
	year = {2017},
	pages = {A97},
}

@article{hammond_linking_2017,
	title = {Linking the {Climate} and {Thermal} {Phase} {Curve} of 55 {Cancri} e},
	volume = {849},
	issn = {0004-637X, 1538-4357},
	url = {https://iopscience.iop.org/article/10.3847/1538-4357/aa9328},
	doi = {10.3847/1538-4357/aa9328},
	language = {en},
	number = {2},
	urldate = {2024-06-09},
	journal = {The Astrophysical Journal},
	author = {Hammond, Mark and T. Pierrehumbert, Raymond},
	month = nov,
	year = {2017},
	pages = {152},
}

@article{bourrier_55_2018,
	title = {The 55 {Cancri} system reassessed},
	volume = {619},
	copyright = {https://www.edpsciences.org/en/authors/copyright-and-licensing},
	issn = {0004-6361, 1432-0746},
	url = {https://www.aanda.org/10.1051/0004-6361/201833154},
	doi = {10.1051/0004-6361/201833154},
	language = {en},
	urldate = {2024-06-20},
	journal = {Astronomy \& Astrophysics},
	author = {Bourrier, V. and Dumusque, X. and Dorn, C. and Henry, G. W. and Astudillo-Defru, N. and Rey, J. and Benneke, B. and Hébrard, G. and Lovis, C. and Demory, B. O. and Moutou, C. and Ehrenreich, D.},
	month = nov,
	year = {2018},
	pages = {A1},
}

@article{li_rovibrational_2015,
	title = {{ROVIBRATIONAL} {LINE} {LISTS} {FOR} {NINE} {ISOTOPOLOGUES} {OF} {THE} {CO} {MOLECULE} {IN} {THE} {X1Σ}+ {GROUND} {ELECTRONIC} {STATE}},
	volume = {216},
	issn = {0067-0049},
	url = {https://dx.doi.org/10.1088/0067-0049/216/1/15},
	doi = {10.1088/0067-0049/216/1/15},
	language = {en},
	number = {1},
	urldate = {2024-06-26},
	journal = {The Astrophysical Journal Supplement Series},
	publisher = {The American Astronomical Society},
	author = {Li, Gang and Gordon, Iouli E. and Rothman, Laurence S. and Tan, Yan and Hu, Shui-Ming and Kassi, Samir and Campargue, Alain and Medvedev, Emile S.},
	month = jan,
	year = {2015},
	pages = {15},
}

@misc{patel_jwst_2024,
	title = {{JWST} reveals a rapid and strong day side variability of 55 {Cancri} e},
	url = {http://arxiv.org/abs/2407.12898},
	language = {en},
	urldate = {2024-07-20},
	publisher = {arXiv},
	author = {Patel, J. A. and Brandeker, A. and Kitzmann, D. and de la Roche, D. J. M. Petit dit and Bello-Arufe, A. and Heng, K. and Valdés, E. Meier and Persson, C. M. and Zhang, M. and Demory, B.-O. and Bourrier, V. and Deline, A. and Ehrenreich, D. and Fridlund, M. and Hu, R. and Lendl, M. and Oza, A. V. and Alibert, Y. and Hooton, M. J.},
	month = jul,
	year = {2024},
	note = {arXiv:2407.12898 [astro-ph]},
}

@article{hu_secondary_2024,
	title = {A secondary atmosphere on the rocky exoplanet 55 {Cancri} e},
	volume = {630},
	issn = {0028-0836, 1476-4687},
	url = {https://www.nature.com/articles/s41586-024-07432-x},
	doi = {10.1038/s41586-024-07432-x},
	language = {en},
	number = {8017},
	urldate = {2024-08-03},
	journal = {Nature},
	author = {Hu, Renyu and Bello-Arufe, Aaron and Zhang, Michael and Paragas, Kimberly and Zilinskas, Mantas and Van Buchem, Christiaan and Bess, Michael and Patel, Jayshil and Ito, Yuichi and Damiano, Mario and Scheucher, Markus and Oza, Apurva V. and Knutson, Heather A. and Miguel, Yamila and Dragomir, Diana and Brandeker, Alexis and Demory, Brice-Olivier},
	month = jun,
	year = {2024},
	pages = {609--612},
}

@article{zhan_novel_2024,
	title = {Novel {Atmospheric} {Dynamics} {Shape} the {Inner} {Edge} of the {Habitable} {Zone} around {White} {Dwarfs}},
	volume = {971},
	copyright = {All rights reserved},
	issn = {0004-637X},
	url = {https://dx.doi.org/10.3847/1538-4357/ad54c1},
	doi = {10.3847/1538-4357/ad54c1},
	language = {en},
	number = {2},
	urldate = {2024-08-23},
	journal = {The Astrophysical Journal},
	publisher = {The American Astronomical Society},
	author = {Zhan, Ruizhi and Koll, Daniel D. B. and Ding, Feng},
	month = aug,
	year = {2024},
	pages = {125},
}

@article{heng_transient_2023,
	title = {The {Transient} {Outgassed} {Atmosphere} of 55 {Cancri} e},
	volume = {956},
	issn = {2041-8205},
	url = {https://dx.doi.org/10.3847/2041-8213/acfe05},
	doi = {10.3847/2041-8213/acfe05},
	language = {en},
	number = {1},
	urldate = {2024-09-02},
	journal = {The Astrophysical Journal Letters},
	publisher = {The American Astronomical Society},
	author = {Heng, Kevin},
	month = oct,
	year = {2023},
	pages = {L20},
}

@article{tsiaras_detection_2016,
	title = {{DETECTION} {OF} {AN} {ATMOSPHERE} {AROUND} {THE} {SUPER}-{EARTH} 55 {CANCRI} {E}},
	volume = {820},
	issn = {0004-637X},
	url = {https://dx.doi.org/10.3847/0004-637X/820/2/99},
	doi = {10.3847/0004-637X/820/2/99},
	language = {en},
	number = {2},
	urldate = {2024-09-02},
	journal = {The Astrophysical Journal},
	publisher = {The American Astronomical Society},
	author = {Tsiaras, A. and Rocchetto, M. and Waldmann, I. P. and Venot, O. and Varley, R. and Morello, G. and Damiano, M. and Tinetti, G. and Barton, E. J. and Yurchenko, S. N. and Tennyson, J.},
	month = mar,
	year = {2016},
	pages = {99},
}

@article{angelo_case_2017,
	title = {A {Case} for an {Atmosphere} on {Super}-{Earth} 55 {Cancri} e},
	volume = {154},
	issn = {1538-3881},
	url = {https://dx.doi.org/10.3847/1538-3881/aa9278},
	doi = {10.3847/1538-3881/aa9278},
	language = {en},
	number = {6},
	urldate = {2024-09-02},
	journal = {The Astronomical Journal},
	publisher = {The American Astronomical Society},
	author = {Angelo, Isabel and Hu, Renyu},
	month = nov,
	year = {2017},
	pages = {232},
}

@article{jindal_characterization_2020,
	title = {Characterization of the {Atmosphere} of {Super}-{Earth} 55 {Cancri} e {Using} {High}-resolution {Ground}-based {Spectroscopy}},
	volume = {160},
	issn = {1538-3881},
	url = {https://dx.doi.org/10.3847/1538-3881/aba1eb},
	doi = {10.3847/1538-3881/aba1eb},
	language = {en},
	number = {3},
	urldate = {2024-09-02},
	journal = {The Astronomical Journal},
	publisher = {The American Astronomical Society},
	author = {Jindal, Abhinav and Mooij, Ernst J. W. de and Jayawardhana, Ray and Deibert, Emily K. and Brogi, Matteo and Rustamkulov, Zafar and Fortney, Jonathan J. and Hood, Callie E. and Morley, Caroline V.},
	month = aug,
	year = {2020},
	pages = {101},
}

@article{mercier_revisiting_2022,
	title = {Revisiting the {Iconic} {Spitzer} {Phase} {Curve} of 55 {Cancri} e: {Hotter} {Dayside}, {Cooler} {Nightside}, and {Smaller} {Phase} {Offset}},
	volume = {164},
	issn = {1538-3881},
	shorttitle = {Revisiting the {Iconic} {Spitzer} {Phase} {Curve} of 55 {Cancri} e},
	url = {https://dx.doi.org/10.3847/1538-3881/ac8f22},
	doi = {10.3847/1538-3881/ac8f22},
	language = {en},
	number = {5},
	urldate = {2024-09-02},
	journal = {The Astronomical Journal},
	publisher = {The American Astronomical Society},
	author = {Mercier, Samson J. and Dang, Lisa and Gass, Alexander and Cowan, Nicolas B. and Bell, Taylor J.},
	month = oct,
	year = {2022},
	pages = {204},
}

@article{vallis_isca_2018,
	title = {Isca, v1.0: a framework for the global modelling of the atmospheres of {Earth} and other planets at varying levels of complexity},
	volume = {11},
	issn = {1991-959X},
	shorttitle = {Isca, v1.0},
	url = {https://gmd.copernicus.org/articles/11/843/2018/gmd-11-843-2018.html},
	doi = {10.5194/gmd-11-843-2018},
	language = {English},
	number = {3},
	urldate = {2024-09-09},
	journal = {Geoscientific Model Development},
	publisher = {Copernicus GmbH},
	author = {Vallis, Geoffrey K. and Colyer, Greg and Geen, Ruth and Gerber, Edwin and Jucker, Martin and Maher, Penelope and Paterson, Alexander and Pietschnig, Marianne and Penn, James and Thomson, Stephen I.},
	month = mar,
	year = {2018},
	pages = {843--859},
}

@article{demory_map_2016,
	title = {A map of the large day–night temperature gradient of a super-{Earth} exoplanet},
	volume = {532},
	copyright = {2016 Springer Nature Limited},
	issn = {1476-4687},
	url = {https://www.nature.com/articles/nature17169},
	doi = {10.1038/nature17169},
	language = {en},
	number = {7598},
	urldate = {2024-10-04},
	journal = {Nature},
	publisher = {Nature Publishing Group},
	author = {Demory, Brice-Olivier and Gillon, Michael and de Wit, Julien and Madhusudhan, Nikku and Bolmont, Emeline and Heng, Kevin and Kataria, Tiffany and Lewis, Nikole and Hu, Renyu and Krick, Jessica and Stamenković, Vlada and Benneke, Björn and Kane, Stephen and Queloz, Didier},
	month = apr,
	year = {2016},
	pages = {207--209},
}

@article{demory_variability_2016,
	title = {Variability in the super-{Earth} 55 {Cnc} e},
	volume = {455},
	issn = {0035-8711},
	url = {https://doi.org/10.1093/mnras/stv2239},
	doi = {10.1093/mnras/stv2239},
	number = {2},
	urldate = {2024-10-04},
	journal = {Monthly Notices of the Royal Astronomical Society},
	author = {Demory, Brice-Olivier and Gillon, Michael and Madhusudhan, Nikku and Queloz, Didier},
	month = jan,
	year = {2016},
	pages = {2018--2027},
}

@article{valdes_weak_2022,
	title = {Weak evidence for variable occultation depth of 55 {Cnc} e with {TESS}},
	volume = {663},
	copyright = {© E. A. Meier Valdés et al. 2022},
	issn = {0004-6361, 1432-0746},
	url = {https://www.aanda.org/articles/aa/abs/2022/07/aa43768-22/aa43768-22.html},
	doi = {10.1051/0004-6361/202243768},
	language = {en},
	urldate = {2024-10-04},
	journal = {Astronomy \& Astrophysics},
	publisher = {EDP Sciences},
	author = {Valdés, E. A. Meier and Morris, B. M. and Wells, R. D. and Schanche, N. and Demory, B.-O.},
	month = jul,
	year = {2022},
	pages = {A95},
}

@misc{loftus_extreme_2024,
	title = {Extreme {Weather} {Variability} on {Hot} {Rocky} {Exoplanet} 55 {Cancri} e {Explained} by {Magma} {Temperature}-{Cloud} {Feedback}},
	url = {http://arxiv.org/abs/2409.16270},
	language = {en},
	urldate = {2024-10-15},
	publisher = {arXiv},
	author = {Loftus, Kaitlyn and Luo, Yangcheng and Fan, Bowen and Kite, Edwin S.},
	month = sep,
	year = {2024},
	note = {arXiv:2409.16270 [astro-ph]},
}

@article{demory_55_2023,
	title = {55 {Cancri} e’s occultation captured with {CHEOPS}},
	volume = {669},
	copyright = {https://creativecommons.org/licenses/by/4.0},
	issn = {0004-6361, 1432-0746},
	url = {https://www.aanda.org/10.1051/0004-6361/202244894},
	doi = {10.1051/0004-6361/202244894},
	language = {en},
	urldate = {2024-10-15},
	journal = {Astronomy \& Astrophysics},
	author = {Demory, B.-O. and Sulis, S. and Meier Valdés, E. and Delrez, L. and Brandeker, A. and Billot, N. and Fortier, A. and Hoyer, S. and Sousa, S. G. and Heng, K. and Lendl, M. and Krenn, A. and Morris, B. M. and Patel, J. A. and Alibert, Y. and Alonso, R. and Anglada, G. and Bárczy, T. and Barrado, D. and Barros, S. C. C. and Baumjohann, W. and Beck, M. and Beck, T. and Benz, W. and Bonfils, X. and Broeg, C. and Buder, M. and Cabrera, J. and Charnoz, S. and Collier Cameron, A. and Cottard, H. and Csizmadia, Sz. and Davies, M. B. and Deleuil, M. and Demangeon, O. D. S. and Ehrenreich, D. and Erikson, A. and Fossati, L. and Fridlund, M. and Gandolfi, D. and Gillon, M. and Güdel, M. and Isaak, K. G. and Kiss, L. L. and Laskar, J. and Lecavelier Des Etangs, A. and Lovis, C. and Luntzer, A. and Magrin, D. and Marafatto, L. and Maxted, P. F. L. and Nascimbeni, V. and Olofsson, G. and Ottensamer, R. and Pagano, I. and Pallé, E. and Peter, G. and Piotto, G. and Pollacco, D. and Queloz, D. and Ragazzoni, R. and Rando, N. and Ratti, F. and Rauer, H. and Ribas, I. and Santos, N. C. and Scandariato, G. and Ségransan, D. and Simon, A. E. and Smith, A. M. S. and Steller, M. and Szabó, Gy. M. and Thomas, N. and Udry, S. and Van Grootel, V. and Walton, N. A.},
	month = jan,
	year = {2023},
	pages = {A64},
}

@article{yurchenko_exomol_2022,
	title = {{ExoMol} line lists – {XLIV}. {Infrared} and ultraviolet line list for silicon monoxide ({28Si16O})},
	volume = {510},
	issn = {0035-8711},
	url = {https://doi.org/10.1093/mnras/stab3267},
	doi = {10.1093/mnras/stab3267},
	number = {1},
	urldate = {2024-10-18},
	journal = {Monthly Notices of the Royal Astronomical Society},
	author = {Yurchenko, Sergei N and Tennyson, Jonathan and Syme, Anna-Maree and Adam, Ahmad Y and Clark, Victoria H J and Cooper, Bridgette and Dobney, C Pria and Donnelly, Shaun T E and Gorman, Maire N and Lynas-Gray, Anthony E and Meltzer, Thomas and Owens, Alec and Qu, Qianwei and Semenov, Mikhail and Somogyi, Wilfrid and Upadhyay, Apoorva and Wright, Samuel and Zapata Trujillo, Juan C},
	month = feb,
	year = {2022},
	pages = {903--919},
}

@article{meier_interior_2023,
	title = {Interior dynamics of super-{Earth} 55 {Cancri} e},
	volume = {678},
	copyright = {https://creativecommons.org/licenses/by/4.0},
	issn = {0004-6361, 1432-0746},
	url = {https://www.aanda.org/10.1051/0004-6361/202346950},
	doi = {10.1051/0004-6361/202346950},
	language = {en},
	urldate = {2024-11-07},
	journal = {Astronomy \& Astrophysics},
	author = {Meier, Tobias G. and Bower, Dan J. and Lichtenberg, Tim and Hammond, Mark and Tackley, Paul J.},
	month = oct,
	year = {2023},
	pages = {A29},
}

@article{deibert_near-infrared_2021,
	title = {A {Near}-infrared {Chemical} {Inventory} of the {Atmosphere} of 55 {Cancri} e},
	volume = {161},
	issn = {0004-6256, 1538-3881},
	url = {https://iopscience.iop.org/article/10.3847/1538-3881/abe768},
	doi = {10.3847/1538-3881/abe768},
	language = {en},
	number = {5},
	urldate = {2024-11-07},
	journal = {The Astronomical Journal},
	author = {Deibert, Emily K. and De Mooij, Ernst J. W. and Jayawardhana, Ray and Ridden-Harper, Andrew and Sivanandam, Suresh and Karjalainen, Raine and Karjalainen, Marie},
	month = may,
	year = {2021},
	pages = {209},
}

@article{rothman_energy_1992,
	series = {Special {Issue} {Conference} on {Molecular} {Spectroscopic} {Databases}},
	title = {Energy levels, intensities, and linewidths of atmospheric carbon dioxide bands},
	volume = {48},
	issn = {0022-4073},
	url = {https://www.sciencedirect.com/science/article/pii/002240739290119O},
	doi = {10.1016/0022-4073(92)90119-O},
	number = {5},
	urldate = {2024-12-19},
	journal = {Journal of Quantitative Spectroscopy and Radiative Transfer},
	author = {Rothman, L. S. and Hawkins, R. L. and Wattson, R. B. and Gamache, R. R.},
	month = nov,
	year = {1992},
	pages = {537--566},
}

@article{karman_update_2019,
	title = {Update of the {HITRAN} collision-induced absorption section},
	volume = {328},
	issn = {0019-1035},
	url = {https://www.sciencedirect.com/science/article/pii/S0019103518306997},
	doi = {10.1016/j.icarus.2019.02.034},
	urldate = {2024-12-19},
	journal = {Icarus},
	author = {Karman, Tijs and Gordon, Iouli E. and van der Avoird, Ad and Baranov, Yury I. and Boulet, Christian and Drouin, Brian J. and Groenenboom, Gerrit C. and Gustafsson, Magnus and Hartmann, Jean-Michel and Kurucz, Robert L. and Rothman, Laurence S. and Sun, Kang and Sung, Keeyoon and Thalman, Ryan and Tran, Ha and Wishnow, Edward H. and Wordsworth, Robin and Vigasin, Andrey A. and Volkamer, Rainer and van der Zande, Wim J.},
	month = aug,
	year = {2019},
	pages = {160--175},
}

@article{faure_pressure_2013,
	title = {Pressure broadening of water and carbon monoxide transitions by molecular hydrogen at high temperatures},
	volume = {116},
	issn = {0022-4073},
	url = {https://www.sciencedirect.com/science/article/pii/S0022407312004268},
	doi = {10.1016/j.jqsrt.2012.09.015},
	urldate = {2024-12-19},
	journal = {Journal of Quantitative Spectroscopy and Radiative Transfer},
	author = {Faure, A. and Wiesenfeld, L. and Drouin, B. J. and Tennyson, J.},
	month = feb,
	year = {2013},
	pages = {79--86},
}

@article{gordon_hitran2016_2017,
	series = {{HITRAN2016} {Special} {Issue}},
	title = {The {HITRAN2016} molecular spectroscopic database},
	volume = {203},
	issn = {0022-4073},
	url = {https://www.sciencedirect.com/science/article/pii/S0022407317301073},
	doi = {10.1016/j.jqsrt.2017.06.038},
	urldate = {2024-12-19},
	journal = {Journal of Quantitative Spectroscopy and Radiative Transfer},
	author = {Gordon, I. E. and Rothman, L. S. and Hill, C. and Kochanov, R. V. and Tan, Y. and Bernath, P. F. and Birk, M. and Boudon, V. and Campargue, A. and Chance, K. V. and Drouin, B. J. and Flaud, J. -M. and Gamache, R. R. and Hodges, J. T. and Jacquemart, D. and Perevalov, V. I. and Perrin, A. and Shine, K. P. and Smith, M. -A. H. and Tennyson, J. and Toon, G. C. and Tran, H. and Tyuterev, V. G. and Barbe, A. and Császár, A. G. and Devi, V. M. and Furtenbacher, T. and Harrison, J. J. and Hartmann, J. -M. and Jolly, A. and Johnson, T. J. and Karman, T. and Kleiner, I. and Kyuberis, A. A. and Loos, J. and Lyulin, O. M. and Massie, S. T. and Mikhailenko, S. N. and Moazzen-Ahmadi, N. and Müller, H. S. P. and Naumenko, O. V. and Nikitin, A. V. and Polyansky, O. L. and Rey, M. and Rotger, M. and Sharpe, S. W. and Sung, K. and Starikova, E. and Tashkun, S. A. and Auwera, J. Vander and Wagner, G. and Wilzewski, J. and Wcisło, P. and Yu, S. and Zak, E. J.},
	month = dec,
	year = {2017},
	pages = {3--69},
}

@article{guest_predicting_2024,
	title = {Predicting the rotational dependence of line broadening using machine learning},
	volume = {401},
	issn = {0022-2852},
	url = {https://www.sciencedirect.com/science/article/pii/S0022285224000286},
	doi = {10.1016/j.jms.2024.111901},
	urldate = {2024-12-19},
	journal = {Journal of Molecular Spectroscopy},
	author = {Guest, Elizabeth R. and Tennyson, Jonathan and Yurchenko, Sergei N.},
	month = mar,
	year = {2024},
	pages = {111901},
}

@article{somogyi_calculation_2021,
	title = {Calculation of electric quadrupole linestrengths for diatomic molecules: {Application} to the {H2}, {CO}, {HF}, and {O2} molecules},
	volume = {155},
	issn = {0021-9606},
	shorttitle = {Calculation of electric quadrupole linestrengths for diatomic molecules},
	url = {https://doi.org/10.1063/5.0063256},
	doi = {10.1063/5.0063256},
	number = {21},
	urldate = {2024-12-19},
	journal = {The Journal of Chemical Physics},
	author = {Somogyi, W. and Yurchenko, S. N. and Yachmenev, A.},
	month = dec,
	year = {2021},
	pages = {214303},
}

@article{shemansky_n2_1969,
	title = {N2 {Vegard}–{Kaplan} {System} in {Absorption}},
	volume = {51},
	issn = {0021-9606},
	url = {https://doi.org/10.1063/1.1672058},
	doi = {10.1063/1.1672058},
	number = {2},
	urldate = {2024-12-19},
	journal = {The Journal of Chemical Physics},
	author = {Shemansky, D. E.},
	month = jul,
	year = {1969},
	pages = {689--700},
}

@article{western_spectrum_2018,
	title = {The spectrum of {N2} from 4,500 to 15,700 cm−1 revisited with pgopher},
	volume = {219},
	issn = {0022-4073},
	url = {https://www.sciencedirect.com/science/article/pii/S0022407318304497},
	doi = {10.1016/j.jqsrt.2018.07.017},
	urldate = {2024-12-19},
	journal = {Journal of Quantitative Spectroscopy and Radiative Transfer},
	author = {Western, Colin M. and Carter-Blatchford, Luke and Crozet, Patrick and Ross, Amanda J. and Morville, Jérôme and Tokaryk, Dennis W.},
	month = nov,
	year = {2018},
	pages = {127--141},
}

@article{western_pgopher_2017,
	series = {Satellite {Remote} {Sensing} and {Spectroscopy}: {Joint} {ACE}-{Odin} {Meeting}, {October} 2015},
	title = {{PGOPHER}: {A} program for simulating rotational, vibrational and electronic spectra},
	volume = {186},
	issn = {0022-4073},
	shorttitle = {{PGOPHER}},
	url = {https://www.sciencedirect.com/science/article/pii/S0022407316300437},
	doi = {10.1016/j.jqsrt.2016.04.010},
	urldate = {2024-12-19},
	journal = {Journal of Quantitative Spectroscopy and Radiative Transfer},
	author = {Western, Colin M.},
	month = jan,
	year = {2017},
	pages = {221--242},
}

@article{jans_rovibronic_2024,
	title = {Rovibronic molecular line list for the {N2}({C3Πu}−{B3Πg}){\textless}math{\textgreater}{\textless}mrow is="true"{\textgreater}{\textless}msub is="true"{\textgreater}{\textless}mrow is="true"{\textgreater}{\textless}/mrow{\textgreater}{\textless}mrow is="true"{\textgreater}{\textless}mn is="true"{\textgreater}2{\textless}/mn{\textgreater}{\textless}/mrow{\textgreater}{\textless}/msub{\textgreater}{\textless}mrow is="true"{\textgreater}{\textless}mo is="true"{\textgreater}({\textless}/mo{\textgreater}{\textless}msup is="true"{\textgreater}{\textless}mrow is="true"{\textgreater}{\textless}mi is="true"{\textgreater}{C}{\textless}/mi{\textgreater}{\textless}/mrow{\textgreater}{\textless}mrow is="true"{\textgreater}{\textless}mn is="true"{\textgreater}3{\textless}/mn{\textgreater}{\textless}/mrow{\textgreater}{\textless}/msup{\textgreater}{\textless}msub is="true"{\textgreater}{\textless}mrow is="true"{\textgreater}{\textless}mi is="true"{\textgreater}Π{\textless}/mi{\textgreater}{\textless}/mrow{\textgreater}{\textless}mrow is="true"{\textgreater}{\textless}mi is="true"{\textgreater}u{\textless}/mi{\textgreater}{\textless}/mrow{\textgreater}{\textless}/msub{\textgreater}{\textless}mo is="true"{\textgreater}−{\textless}/mo{\textgreater}{\textless}msup is="true"{\textgreater}{\textless}mrow is="true"{\textgreater}{\textless}mi is="true"{\textgreater}{B}{\textless}/mi{\textgreater}{\textless}/mrow{\textgreater}{\textless}mrow is="true"{\textgreater}{\textless}mn is="true"{\textgreater}3{\textless}/mn{\textgreater}{\textless}/mrow{\textgreater}{\textless}/msup{\textgreater}{\textless}msub is="true"{\textgreater}{\textless}mrow is="true"{\textgreater}{\textless}mi is="true"{\textgreater}Π{\textless}/mi{\textgreater}{\textless}/mrow{\textgreater}{\textless}mrow is="true"{\textgreater}{\textless}mi is="true"{\textgreater}g{\textless}/mi{\textgreater}{\textless}/mrow{\textgreater}{\textless}/msub{\textgreater}{\textless}mo is="true"{\textgreater}){\textless}/mo{\textgreater}{\textless}/mrow{\textgreater}{\textless}/mrow{\textgreater}{\textless}/math{\textgreater} second positive system},
	volume = {312},
	issn = {0022-4073},
	url = {https://www.sciencedirect.com/science/article/pii/S0022407323003278},
	doi = {10.1016/j.jqsrt.2023.108809},
	urldate = {2024-12-19},
	journal = {Journal of Quantitative Spectroscopy and Radiative Transfer},
	author = {Jans, E. R.},
	month = jan,
	year = {2024},
	pages = {108809},
}

@article{gamache_thermodynamic_2023,
	title = {Thermodynamic {Functions} for {N2} from the {Total} {Partition} {Sum} and its {Moments}},
	volume = {52},
	issn = {0047-2689},
	url = {https://doi.org/10.1063/5.0137083},
	doi = {10.1063/5.0137083},
	number = {2},
	urldate = {2024-12-19},
	journal = {Journal of Physical and Chemical Reference Data},
	author = {Gamache, Robert R. and Orphanos, Nicholas G.},
	month = may,
	year = {2023},
	pages = {023101},
}

@article{souza_photoabsorption_1994,
	title = {Photoabsorption {Cross} of {Ar}, {N}$_{\textrm{2}}$ and {Si}({CH}$_{\textrm{3}}$ )$_{\textrm{4}}$ {Derived} from {Electron} {Energy} {Loss} {Spectroscopy}},
	volume = {5},
	issn = {0103-5053},
	url = {https://jbcs.sbq.org.br/audiencia_pdf.asp?aid2=140&nomeArquivo=v5n2-01.pdf},
	doi = {10.5935/0103-5053.19940010},
	language = {en},
	number = {2},
	urldate = {2024-12-19},
	journal = {Journal Of The Brazilian Chemical Society},
	author = {Souza, A.C.A. and Srivastava, S.K.},
	year = {1994},
	pages = {59--65},
}

@article{thomson_hierarchical_2019,
	title = {Hierarchical {Modeling} of {Solar} {System} {Planets} with {Isca}},
	volume = {10},
	copyright = {http://creativecommons.org/licenses/by/3.0/},
	issn = {2073-4433},
	url = {https://www.mdpi.com/2073-4433/10/12/803},
	doi = {10.3390/atmos10120803},
	language = {en},
	number = {12},
	urldate = {2024-12-19},
	journal = {Atmosphere},
	publisher = {Multidisciplinary Digital Publishing Institute},
	author = {Thomson, Stephen I. and Vallis, Geoffrey K.},
	month = dec,
	year = {2019},
	note = {Number: 12},
	pages = {803},
}

@article{lewis_dependence_2021,
	title = {The dependence of global super-rotation on planetary rotation rate},
	volume = {78},
	issn = {0022-4928, 1520-0469},
	url = {http://arxiv.org/abs/2004.08414},
	doi = {10.1175/JAS-D-20-0326.1},
	number = {4},
	urldate = {2024-12-19},
	journal = {Journal of the Atmospheric Sciences},
	author = {Lewis, Neil T. and Colyer, Greg J. and Read, Peter L.},
	month = apr,
	year = {2021},
	note = {arXiv:2004.08414 [astro-ph]},
	pages = {1245--1258},
}

@article{keller-rudek_mpi-mainz_2013,
	title = {The {MPI}-{Mainz} {UV}/{VIS} {Spectral} {Atlas} of {Gaseous} {Molecules} of {Atmospheric} {Interest}},
	volume = {5},
	issn = {1866-3508},
	url = {https://essd.copernicus.org/articles/5/365/2013/},
	doi = {10.5194/essd-5-365-2013},
	language = {English},
	number = {2},
	urldate = {2024-12-19},
	journal = {Earth System Science Data},
	publisher = {Copernicus GmbH},
	author = {Keller-Rudek, H. and Moortgat, G. K. and Sander, R. and Sörensen, R.},
	month = dec,
	year = {2013},
	pages = {365--373},
}

@article{amundsen_uk_2016,
	title = {The {UK} {Met} {Office} global circulation model with a sophisticated radiation scheme applied to the hot {Jupiter} {HD} 209458b},
	volume = {595},
	copyright = {© ESO, 2016},
	issn = {0004-6361, 1432-0746},
	url = {https://www.aanda.org/articles/aa/abs/2016/11/aa29183-16/aa29183-16.html},
	doi = {10.1051/0004-6361/201629183},
	language = {en},
	urldate = {2024-12-20},
	journal = {Astronomy \& Astrophysics},
	publisher = {EDP Sciences},
	author = {Amundsen, David S. and Mayne, Nathan J. and Baraffe, Isabelle and Manners, James and Tremblin, Pascal and Drummond, Benjamin and Smith, Chris and Acreman, David M. and Homeier, Derek},
	month = nov,
	year = {2016},
	pages = {A36},
}

@article{way_resolving_2017,
	title = {Resolving {Orbital} and {Climate} {Keys} of {Earth} and {Extraterrestrial} {Environments} with {Dynamics} ({ROCKE}-{3D}) 1.0: {A} {General} {Circulation} {Model} for {Simulating} the {Climates} of {Rocky} {Planets}},
	volume = {231},
	issn = {0067-0049},
	shorttitle = {Resolving {Orbital} and {Climate} {Keys} of {Earth} and {Extraterrestrial} {Environments} with {Dynamics} ({ROCKE}-{3D}) 1.0},
	url = {https://dx.doi.org/10.3847/1538-4365/aa7a06},
	doi = {10.3847/1538-4365/aa7a06},
	language = {en},
	number = {1},
	urldate = {2024-12-20},
	journal = {The Astrophysical Journal Supplement Series},
	publisher = {The American Astronomical Society},
	author = {Way, M. J. and Aleinov, I. and Amundsen, David S. and Chandler, M. A. and Clune, T. L. and Genio, A. D. Del and Fujii, Y. and Kelley, M. and Kiang, N. Y. and Sohl, L. and Tsigaridis, K.},
	month = jul,
	year = {2017},
	pages = {12},
}

@article{christie_impact_2021,
	title = {The impact of mixing treatments on cloud modelling in {3D} simulations of hot {Jupiters}},
	volume = {506},
	copyright = {https://creativecommons.org/licenses/by/4.0/},
	issn = {0035-8711, 1365-2966},
	url = {https://academic.oup.com/mnras/article/506/3/4500/6322843},
	doi = {10.1093/mnras/stab2027},
	language = {en},
	number = {3},
	urldate = {2024-12-20},
	journal = {Monthly Notices of the Royal Astronomical Society},
	author = {Christie, D A and Mayne, N J and Lines, S and Parmentier, V and Manners, J and Boutle, I and Drummond, B and Mikal-Evans, T and Sing, D K and Kohary, K},
	month = aug,
	year = {2021},
	pages = {4500--4515},
}

@article{guzewich_3d_2021,
	title = {{3D} {Simulations} of the {Early} {Martian} {Hydrological} {Cycle} {Mediated} by a {H2}-{CO2} {Greenhouse}},
	volume = {126},
	issn = {2169-9100},
	url = {https://onlinelibrary.wiley.com/doi/abs/10.1029/2021JE006825},
	doi = {10.1029/2021JE006825},
	language = {en},
	number = {7},
	urldate = {2024-12-20},
	journal = {Journal of Geophysical Research: Planets},
	author = {Guzewich, Scott D. and Way, Michael J. and Aleinov, Igor and Wolf, Eric T. and Del Genio, Anthony and Wordsworth, Robin and Tsigaridis, Kostas},
	year = {2021},
	note = {\_eprint: https://onlinelibrary.wiley.com/doi/pdf/10.1029/2021JE006825},
	pages = {e2021JE006825},
}

@misc{zilinskas_characterising_2025,
	title = {Characterising the {Atmosphere} of 55 {Cancri} e: {1D} {Forward} {Model} {Grid} for {Current} and {Future} {JWST} {Observations}},
	shorttitle = {Characterising the {Atmosphere} of 55 {Cancri} e},
	url = {http://arxiv.org/abs/2503.15844},
	doi = {10.48550/arXiv.2503.15844},
	urldate = {2025-03-21},
	publisher = {arXiv},
	author = {Zilinskas, Mantas and Buchem, Christiaan van and Zieba, Sebastian and Miguel, Yamila and Sandford, Emily and Hu, Renyu and Patel, Jayshil and Bello-Arufe, Aaron and Janssen, Leoni and Tsai, Shang-Min and Dragomir, Diana and Zhang, Michael},
	month = mar,
	year = {2025},
	note = {arXiv:2503.15844 [astro-ph]},
}

@misc{ji_cosmic_2025,
	title = {The {Cosmic} {Shoreline} {Revisited}: {A} {Metric} for {Atmospheric} {Retention} {Informed} by {Hydrodynamic} {Escape}},
	shorttitle = {The {Cosmic} {Shoreline} {Revisited}},
	url = {http://arxiv.org/abs/2504.19872},
	doi = {10.48550/arXiv.2504.19872},
	urldate = {2025-05-03},
	publisher = {arXiv},
	author = {Ji, Xuan and Chatterjee, Richard D. and Coy, Brandon Park and Kite, Edwin S.},
	month = apr,
	year = {2025},
	note = {arXiv:2504.19872 [astro-ph]},
}

@article{tennyson_exomol_2018,
	title = {The {ExoMol} {Atlas} of {Molecular} {Opacities}},
	volume = {6},
	issn = {2218-2004},
	url = {http://arxiv.org/abs/1805.03711},
	doi = {10.3390/atoms6020026},
	number = {2},
	urldate = {2025-05-27},
	journal = {Atoms},
	author = {Tennyson, Jonathan and Yurchenko, Sergei N.},
	month = may,
	year = {2018},
	note = {arXiv:1805.03711 [astro-ph]},
	pages = {26},
}

@article{polyansky_exomol_2018,
	title = {{ExoMol} molecular line lists {XXX}: a complete high-accuracy line list for water},
	volume = {480},
	issn = {0035-8711},
	shorttitle = {{ExoMol} molecular line lists {XXX}},
	url = {https://doi.org/10.1093/mnras/sty1877},
	doi = {10.1093/mnras/sty1877},
	number = {2},
	urldate = {2025-10-04},
	journal = {Monthly Notices of the Royal Astronomical Society},
	author = {Polyansky, Oleg L and Kyuberis, Aleksandra A and Zobov, Nikolai F and Tennyson, Jonathan and Yurchenko, Sergei N and Lodi, Lorenzo},
	month = oct,
	year = {2018},
	pages = {2597--2608},
}

@article{fegley_volatile_2020,
	title = {Volatile element chemistry during accretion of the earth},
	volume = {80},
	issn = {00092819},
	url = {https://linkinghub.elsevier.com/retrieve/pii/S1611586419300423},
	doi = {10.1016/j.chemer.2019.125594},
	language = {en},
	number = {1},
	urldate = {2025-10-04},
	journal = {Geochemistry},
	author = {Fegley, Bruce and Lodders, Katharina and Jacobson, Nathan S.},
	month = apr,
	year = {2020},
	pages = {125594},
}

@article{crossfield_acme_2012,
	title = {{ACME} stellar spectra: {I}. {Absolutely} calibrated, mostly empirical flux densities of 55 {Cancri} and its transiting planet 55 {Cancri} e⋆},
	volume = {545},
	issn = {0004-6361, 1432-0746},
	shorttitle = {{ACME} stellar spectra},
	url = {http://www.aanda.org/10.1051/0004-6361/201219826},
	doi = {10.1051/0004-6361/201219826},
	language = {en},
	urldate = {2025-11-06},
	journal = {Astronomy \& Astrophysics},
	author = {Crossfield, I. J. M.},
	month = sep,
	year = {2012},
	pages = {A97},
}

@article{yurchenko_exocross_2018,
	title = {{EXOCROSS}: a general program for generating spectra from molecular line lists},
	volume = {614},
	copyright = {http://creativecommons.org/licenses/by/4.0},
	issn = {0004-6361, 1432-0746},
	shorttitle = {{EXOCROSS}},
	url = {https://www.aanda.org/10.1051/0004-6361/201732531},
	doi = {10.1051/0004-6361/201732531},
	language = {en},
	urldate = {2025-12-29},
	journal = {Astronomy \& Astrophysics},
	author = {Yurchenko, Sergei N. and Al-Refaie, Ahmed F. and Tennyson, Jonathan},
	month = jun,
	year = {2018},
	pages = {A131},
}

@article{tian_thermal_2009,
	title = {{THERMAL} {ESCAPE} {FROM} {SUPER} {EARTH} {ATMOSPHERES} {IN} {THE} {HABITABLE} {ZONES} {OF} {M} {STARS}},
	volume = {703},
	issn = {0004-637X},
	url = {https://doi.org/10.1088/0004-637X/703/1/905},
	doi = {10.1088/0004-637X/703/1/905},
	language = {en},
	number = {1},
	urldate = {2026-01-12},
	journal = {The Astrophysical Journal},
	publisher = {The American Astronomical Society},
	author = {Tian, Feng},
	month = sep,
	year = {2009},
	pages = {905},
}

@misc{farhat_magma_2026,
	title = {Magma {Ocean} {Waves} and {Thermal} {Variability} on {Lava} {Worlds}},
	url = {http://arxiv.org/abs/2601.07080},
	doi = {10.48550/arXiv.2601.07080},
	urldate = {2026-01-13},
	publisher = {arXiv},
	author = {Farhat, Mohammad and Chiang, Eugene},
	month = jan,
	year = {2026},
	note = {arXiv:2601.07080 [astro-ph]},
}

@inproceedings{allard_phoenix_2016,
	address = {Lyon, France},
	title = {The {PHOENIX} {Model} {Atmosphere} {Grid} for {Stars}},
	booktitle = {{SF2A}-2016: {Proceedings} of the {Annual} meeting of the {French} {Society} of {Astronomy} and {Astrophysics}},
	author = {Allard, France},
	editor = {Reylé, C. and Richard, J. and Cambrésy, L. and {others}},
	month = dec,
	year = {2016},
	pages = {223--227},
}

@book{elsasser_heat_1942,
	address = {Cambridge, MA},
	series = {Harvard {Meteorological} {Studies}},
	title = {Heat {Transfer} by {Infrared} {Radiation} in the {Atmosphere}},
	number = {6},
	publisher = {Harvard University, Blue Hill Meteorological Observatory},
	author = {Elsasser, Walter M.},
	year = {1942},
}

@article{zdunkowski_investigation_1980,
	title = {An investigation of the structure of typical two-stream-methods for the calculation of solar fluxes and heating rates in clouds},
	volume = {53},
	journal = {Contributions to Atmospheric Physics},
	author = {Zdunkowski, W. G. and Welch, R. M. and Korb, G.},
	year = {1980},
	pages = {147--166},
}

@article{thomson_effects_2019,
	title = {The effects of gravity on the climate and circulation of a terrestrial planet},
	volume = {145},
	issn = {1477-870X},
	url = {https://onlinelibrary.wiley.com/doi/abs/10.1002/qj.3582},
	doi = {10.1002/qj.3582},
	language = {en},
	number = {723},
	urldate = {2026-01-30},
	journal = {Quarterly Journal of the Royal Meteorological Society},
	author = {Thomson, Stephen I. and Vallis, Geoffrey K.},
	year = {2019},
	note = {\_eprint: https://rmets.onlinelibrary.wiley.com/doi/pdf/10.1002/qj.3582},
	pages = {2627--2640},
}

@article{seager_method_2009,
	title = {{ON} {THE} {METHOD} {TO} {INFER} {AN} {ATMOSPHERE} {ON} {A} {TIDALLY} {LOCKED} {SUPER} {EARTH} {EXOPLANET} {AND} {UPPER} {LIMITS} {TO} {GJ} 876d},
	volume = {703},
	issn = {0004-637X},
	url = {https://iopscience.iop.org/article/10.1088/0004-637X/703/2/1884},
	doi = {10.1088/0004-637X/703/2/1884},
	language = {en},
	number = {2},
	urldate = {2026-04-01},
	journal = {The Astrophysical Journal},
	publisher = {IOP Publishing},
	author = {Seager, S. and Deming, D.},
	month = sep,
	year = {2009},
	pages = {1884},
}

@article{fischer_carbon_2020,
	title = {The carbon content of {Earth} and its core},
	volume = {117},
	issn = {0027-8424, 1091-6490},
	url = {https://pnas.org/doi/full/10.1073/pnas.1919930117},
	doi = {10.1073/pnas.1919930117},
	language = {en},
	number = {16},
	urldate = {2026-04-01},
	journal = {Proceedings of the National Academy of Sciences},
	author = {Fischer, Rebecca A. and Cottrell, Elizabeth and Hauri, Erik and Lee, Kanani K. M. and Le Voyer, Marion},
	month = apr,
	year = {2020},
	pages = {8743--8749},
}

@article{seager_dayside_2005,
	title = {On the {Dayside} {Thermal} {Emission} of {Hot} {Jupiters}},
	volume = {632},
	issn = {0004-637X, 1538-4357},
	url = {https://iopscience.iop.org/article/10.1086/444411},
	doi = {10.1086/444411},
	language = {en},
	number = {2},
	urldate = {2026-04-01},
	journal = {The Astrophysical Journal},
	author = {Seager, S. and Richardson, L. J. and Hansen, B. M. S. and Menou, K. and Cho, J. Y.‐K. and Deming, D.},
	month = oct,
	year = {2005},
	pages = {1122--1131},
}

@article{lecuyer_comparison_2000,
	title = {Comparison of carbon, nitrogen and water budgets on {Venus} and the {Earth}},
	volume = {181},
	issn = {0012-821X},
	url = {https://www.sciencedirect.com/science/article/pii/S0012821X00001953},
	doi = {10.1016/S0012-821X(00)00195-3},
	number = {1},
	urldate = {2026-05-24},
	journal = {Earth and Planetary Science Letters},
	author = {Lécuyer, Christophe and Simon, Laurent and Guyot, François},
	month = aug,
	year = {2000},
	pages = {33--40},
}

@article{avice_noble_2022,
	title = {Noble {Gases} and {Stable} {Isotopes} {Track} the {Origin} and {Early} {Evolution} of the {Venus} {Atmosphere}},
	volume = {218},
	issn = {1572-9672},
	url = {https://doi.org/10.1007/s11214-022-00929-9},
	doi = {10.1007/s11214-022-00929-9},
	language = {en},
	number = {8},
	urldate = {2026-05-24},
	journal = {Space Science Reviews},
	author = {Avice, Guillaume and Parai, Rita and Jacobson, Seth and Labidi, Jabrane and Trainer, Melissa G. and Petkov, Mihail P.},
	month = oct,
	year = {2022},
	pages = {60},
}

@article{halliday_origins_2013,
	title = {The origins of volatiles in the terrestrial planets},
	volume = {105},
	issn = {0016-7037},
	url = {https://www.sciencedirect.com/science/article/pii/S0016703712006680},
	doi = {10.1016/j.gca.2012.11.015},
	urldate = {2026-05-24},
	journal = {Geochimica et Cosmochimica Acta},
	author = {Halliday, Alex N.},
	month = mar,
	year = {2013},
	pages = {146--171},
}

@article{essack_low-albedo_2020,
	title = {Low-albedo {Surfaces} of {Lava} {Worlds}},
	volume = {898},
	issn = {0004-637X},
	url = {https://doi.org/10.3847/1538-4357/ab9cba},
	doi = {10.3847/1538-4357/ab9cba},
	language = {en},
	number = {2},
	urldate = {2026-06-04},
	journal = {The Astrophysical Journal},
	publisher = {The American Astronomical Society},
	author = {Essack, Zahra and Seager, Sara and Pajusalu, Mihkel},
	month = aug,
	year = {2020},
	pages = {160},
}

@article{kipping_detection_2020,
	title = {Detection of the {Occultation} of 55 {Cancri} e with {TESS}},
	volume = {4},
	issn = {2515-5172},
	url = {https://doi.org/10.3847/2515-5172/abbc0f},
	doi = {10.3847/2515-5172/abbc0f},
	language = {en},
	number = {9},
	urldate = {2026-06-04},
	journal = {Research Notes of the AAS},
	publisher = {The American Astronomical Society},
	author = {Kipping, David and Jansen, Tiffany},
	month = sep,
	year = {2020},
	pages = {170},
}
\bibliographystyle{aasjournal}

\end{document}